\documentclass[10pt,conference,compsocconf,letterpaper]{IEEEtran}

\newif\ifcomm
\commfalse

\ifcomm
	\newcommand{\mycomm}[3]{{\footnotesize{{\color{#2} \textbf{[#1: #3]}}}}}
\else
    \newcommand{\mycomm}[3]{}
\fi

\newcommand{\GI}[1]{\mycomm{Gil}{blue}{#1}} 
\newcommand{\MM}[1]{\mycomm{MM}{red}{#1}} 
\newcommand{\RBB}[1]{\mycomm{Ran}{purple}{#1}} 
 
\newcommand{\ran}[1]{\RBB{#1}}

\let\subparagraph\relax
\usepackage[compact]{titlesec}
\titlespacing{\section}{0.6pt}{*0.6}{*0.6}
\titlespacing{\subsection}{0.5pt}{*0.5}{*0.5}
\titlespacing{\subsubsection}{0.5pt}{*0.5}{*0.5}

\def\BibTeX{{\rm B\kern-.05em{\sc i\kern-.025em b}\kern-.08em T\kern-.1667em\lower.7ex\hbox{E}\kern-.125emX}}

\usepackage{cite}
\usepackage{algpseudocode}

\usepackage{graphicx}
\usepackage{comment}

\usepackage{caption}
\usepackage{amsthm}
\usepackage{amsmath}
\usepackage{amssymb}
\usepackage{algorithm}
\usepackage{subfig}
\usepackage{xcolor}
\usepackage{blindtext}
\usepackage{thm-restate}
\algtext*{EndFunction}
\algtext*{EndProcedure}
\algtext*{EndIf}
\algtext*{EndFor}
\algtext*{EndWhile}
\algtext*{EndIf}
\newcommand{\Var}{\mathrm{Var}}

\newcommand{\set}[1]{\left\{#1\right\}}
\newcommand{\brackets}[1]{\left[#1\right]}
\newcommand{\ceil}[1]{ \left\lceil{#1}\right\rceil}
\newcommand{\floor}[1]{ \left\lfloor{#1}\right\rfloor}

\newcommand{\parentheses}[1]{ \left({#1}\right)}

\newcommand{\cdotpa}[1]{\cdot\parentheses{#1}}

\title{Faster and More Accurate Measurement through Additive-Error Counters}

\author{\IEEEauthorblockN{Ran Ben Basat}
	\IEEEauthorblockA{Harvard University
	}
	\and
	\IEEEauthorblockN{Gil Einziger}
	\IEEEauthorblockA{Ben Gurion University
	}
	\and
	\IEEEauthorblockN{Michael Mitzenmacher }
	\IEEEauthorblockA{Harvard University	
	}
	\and
	\IEEEauthorblockN{Shay Vargaftik}
	\IEEEauthorblockA{
		VMware Research
	}
}

\date{}

\newtheorem*{observation*}{Observation}

\begin{document}
\newcommand*{\NINEPAGES}{}
\ifdefined \NINEPAGES
\newcommand*{\TENPAGES}{}
\fi
\maketitle

\begin{abstract}
Counters are a fundamental building block for networking applications such as load balancing, traffic engineering, and intrusion detection, which require estimating flow sizes and identifying heavy hitter flows. 
Existing works suggest replacing counters with shorter multiplicative error \emph{estimators} that improve the accuracy by fitting more of them within a given space.
However, such estimators impose a computational overhead that degrades the measurement throughput.
Instead, we propose \emph{additive} error estimators, which are simpler, faster, and more accurate when used for network measurement.
Our solution is rigorously analyzed and empirically evaluated against several other measurement algorithms on real Internet traces. 
For a given error target, we improve the speed of the uncompressed solutions by $5\times$-$30\times$, and the space by up to $4\times$. 
Compared with existing state-of-the-art estimators, our solution is $ 9\times$-$35\times$ faster while being considerably more accurate.

\end{abstract}

\newcommand{\matrixCellWidth}{5.8cm}

\section{Introduction}



Networking applications such as load balancing~\cite{LoadBalancing}, traffic-engineering~\cite{TrafficEngeneering}, SLA enforcement~\cite{SLA}, and intrusion detection~\cite{IntrusionDetection,IntrusionDetection2} require measurement information such as flow sizes and heavy hitter flows.
Computing this information is challenging due to the limited amount of fast memory and the rapid line rates~\cite{Nitro,RHHH,Brick}. 
Such constraints motivate \emph{approximate} measurements which reduce the overheads at the cost of introducing a provably bounded error~\cite{univmon,CountSketch,CountMinSketch,RandomizedCounterSharing,SketchVisor,SpaceSavings}. 

Accordingly, many measurement algorithms use a small number of "shared" counters for providing estimates for all flow sizes instead of tracking each with a dedicated counter.  
Previous work suggests replacing counters used in these methods with shorter probabilistic counters (a.k.a \emph{estimators}) that approximately count up to large numbers with fewer bits~\cite{SAC,DISCO,CEDAR,ICE-Buckets,CASE,Infocom2019}. Such estimators require less memory than regular counters, allowing more \mbox{to fit within a given amount of space.}

Such estimators have been shown to empirically improve the accuracy on networking workloads at the cost of added complexity and reduced speed~\cite{Infocom2019}.  Approximate measurement algorithms that can benefit from such estimators~\cite{CountSketch,CUSketch,univmon} often require significant per-packet processing to calculate multiple hash values or update sophisticated data structures. Sampling techniques~\cite{RHHH,Nitro} reduce the number of packets that need to be processed, increasing speed at the cost of losing accuracy and requiring more memory. 

Our work provides simple and effective estimator techniques that increase the processing speed \emph{and} reduce the required space.  In particular, we make use of the fact that most sketching and sampling based algorithms yield \emph{additive} errors on the order of $N\epsilon $, where $\epsilon$ is pre-selected constant and $N$ is the size of the total count (in terms of number of packets or bytes).  
Therefore, unlike previous work that provided estimators with a {\em multiplicative} error, we focus on estimators that themselves have an {\em additive} error bound.  
As the combination of an additive-error algorithm with a multiplicative-error estimator results in an additive error solution anyway, we study the potential benefits of additive-error estimators for accuracy and speed.
We provide formal accuracy guarantees for our methods, including examples of practical configurations where our approach improves the accuracy. 
We then evaluate our methods empirically on real network traces, and show that they improve the accuracy compared to the state of the art estimators while being $9\times$-$35\times$ faster. 
Further, for a given error target, we improve the speed and space of the uncompressed solutions by $5\times$-$30\times$ and up to $4\times$ respectively. 


\section{Related Work}

We describe the related work in terms of estimators, sketch algorithms, and cache-based counting algorithms.  We note that this terminology does not appear standard and previous work refer to them as "counters" or "approximate counters" (regardless of whether they are counting one object or many);  we find distinguishing the types of \mbox{algorithms in this way clearer.}

\subsubsection{Estimators}
We use the term \emph{Estimator} to refer to a small approximate counter (e.g., a register), which can approximately represent a large number. An estimator generally works via probabilistic increments;  when an item corresponding to that counter arrives, we flip a coin and add one to the estimator with a certain probability.  The estimator's value is used to derive an approximate estimate for the actual count.  In what follows we refer to a {\em probabilistic increment operation} (or PI) as an operation where the estimator may be increased, and an increment as a case where the estimator is incremented (due to a successful coin flip.).  The estimator value is used to estimate the number of PIs associated with estimator.  Estimators differ from each other by the PI probabilities.  Some estimators work for fixed ranges, while others utilize techniques to dynamically increase the counting \mbox{range (generally at the expense of a larger error).}

The \emph{Approximate Counting}~\cite{ApproximateCounting} algorithm is the first estimator we are aware of, and it inspired a substantial number of follow-on works~\cite{ANLS,ANLSUpscaling,CEDAR,ICE-Buckets,CASE,SAC,DISCO} (that we do \mbox{not discuss here).}




\subsubsection{Sketch Algorithms}
Sketch algorithms for keeping large-scale count information in networks are typically composed of arrays of counters.  When a packet arrives, the algorithm applies multiple hash functions to its flow id, 
mapping the flow to a set of counters.
Examples include the Count Min Sketch (CMS)~\cite{CountMinSketch}, the Count Sketch~\cite{CountSketch},  Spectral Bloom Filter~\cite{SpectralBloom}, and the \mbox{Conservative Update (CU) Sketch~\cite{CUSketch}}.
 CMS utilizes multiple counter arrays, 
 where each has a hash function that associates each flow with a counter.
To increment a flow count in CMS, we apply the hash function of each array to the element and increment the corresponding counter. We estimate the count for a flow by returning the minimal value of all of its relevant counters.  
The CU Sketch optimizes the accuracy of CMS in a simple manner. When we add an item to the CMS, we only increment the corresponding counters whose value is minimal. That is, if we read 3,4,3, and 5 then we only increment the counters that show 3 to 4. This optimization avoids unnecessary increments, giving more accurate estimates.  
However, while CMS supports decrements, the CU Sketch does not. 

\emph{CounterBraids}~\cite{CounterBraids} introduce an hierarchical structure which reduces the average counter length of CMS at the expense of much slower decoding process.  
Alternatively, \emph{Randomized Counter Sharing (RCS)}~\cite{RandomizedCounterSharing} only updates a single randomly selected counter to achieve a faster update time, and sum all counters for an estimate.  
NitroSketch~\cite{Nitro} takes RCS a step further, 
providing several techniques to accelerate software sketches in virtual switches, including geometric sampling.  In general, NitroSketch increases the required space, but accelerates the sketch's throughput in software.
\emph{Counter Tree}~\cite{countertree} introduces multiple \emph{virtual counters} that extend multiple physical counters in a tree structure. 
Counter Tree also trades off speed for space efficiency. 

The use of estimator algorithms to compress sketch counters is particularly relevant to our work.
\emph{Small Active Counters}~\cite{SAC} implement an array of estimators, where each estimator keeps track of an exponent and an estimation part.  The exponent part determines the probability of success for the PI, which increments the estimation part. When the estimation part reaches its maximum value, the exponent increases and the estimation part resets to 0.  
The \emph{DISCO}~\cite{DISCO} algorithm improves~\cite{SAC}'s accuracy and supports weighted updates (where a counter increases by a given quantity). The work of~\cite{ANLSUpscaling} introduces a way to gradually increase the measurement scale when a counter overflows at the expense of larger error. 
\emph{CEDAR}~\cite{CEDAR} proves that their estimation function is optimal for min-max relative error. 
\emph{ICE-Buckets} \cite{ICE-Buckets} uses multiple measurement scales within a single array of estimators to reduce the error, while 
\emph{CASE}~\cite{CASE} shows that using a cache to monitor the largest flows accurately improves the estimation accuracy. Most relevant to our paper, the recent work of~\cite{Infocom2019} suggests a new estimator with multiple counter scales and demonstrates an empirical error reduction at \mbox{the expense of a slower run-time.}

\subsubsection{Cache-Based Algorithms}
We refer to cache-based algorithms for the class of algorithms that maintain a small cache of entries, each containing generally at least the flow identifier and its packet or byte count~\cite{frequent4,SpaceSavingIsTheBest,HashPipe,HeavyHitters,10.14778/3297753.3297762}.  To keep space usage reasonable, cache-based \mbox{algorithms do not keep counts for all flows.}

Cache-based algorithms differ from each other in their cache policy, governing when to admit a new flow and which flow to evict when admitting a new flow to a full cache.  In software deployments, cache-based algorithms often yield an attractive space/accuracy trade-off when compared to sketch algorithms~\cite{SpaceSavingIsTheBest,SpaceSavingIsTheBest2010,SpaceSavingIsTheBest2011}.  The \emph{Misra-Gries (MG)} algorithm~\cite{misra1982finding} is perhaps the most famous cache-based algorithm, and requires logarithmic update time.  The works of~\cite{frequent4,BatchDecrement} independently improve the \mbox{update time to a constant for unweighted streams.}

The Space-Saving algorithm~\cite{SpaceSavings} maintains a cache of flow entries, each with its own packet (or byte) counter. When a packet from an unmonitored flow arrives to a full cache we evict the entry whose packet count $m$ is the smallest among all monitored flows (there may be more than one), and admit the unmonitored flow with an initial packet count of $m+1$. Space saving also supports weighted updates. In that case, we admit a new entry with a count of $m+w$ where $w$ is the weight of the update.  
Formally, when the Space-Saving algorithm is configured with $\epsilon_A^{-1}$ entries (for some $\epsilon_A$ in $(0,1)$), it provides an $N\epsilon_A $ additive error when $N$ is the totoal number of packets. 

The Randomized Admission Policy (RAP)~\cite{RAP} provides a simple heuristic that optimizes cache-based algorithms for heavy-tailed workloads.
RAP leverages the fact that most packets belong to small flows, so admitting them to the cache means that we stop monitoring important flows. Therefore, RAP admits a new flow with probability $\frac{w}{m+w}$ ($\frac{1}{m+1}$ for unweighted streams). The technique gives a significant empirical improvement in accuracy but currently lacks formal correctness proofs. The authors also suggest $d$-way RAP, which has smaller implementation overhead by using limited associativity arrays. They show that 16-way RAP achieves almost the same results as its fully associative counterpart. 

Cache-based algorithms can also process weighted inputs, but generally requires more sophisticated algorithms and resources. The Space-Saving algorithm can be implemented with constant update complexity for unit weights and with a logarithmic complexity for general weights. Recent works suggest weighted cache-based algorithms with a constant update complexity~\cite{dimsum,IMSUM}, \mbox{at the expense of a larger space requirement. }

To the best of our knowledge, estimators were not previously suggested for cache-based algorithms. A possible explanation lies with the data structures associated with counter algorithms. Specifically, flow identifiers are typically 13 bytes long, and such algorithms also have other additional space overheads. When the actual counters are typically 4-8 bytes long the benefit of reducing the counter size is limited. 
We show that estimators can benefit cache-based algorithms, \mbox{especially when optimizing their data structures for space.}



\section{Additive-error Estimator}

We start by presenting our estimator.  
In this section, we assume that the required counting range ($N$) is known in advance. 
We later show in Section~\ref{sec:dynamicN} how to dynamically increase the counting range. 
Our additive error estimator can count up to $N$ with an \emph{additive} error of at most $N\epsilon$, with probability at least $1-\delta$. 
We emphasize again that additive guarantees are uncommon in estimator algorithms, which typically provide \emph{multiplicative} error~\cite{ICE-Buckets,CASE,ANLSUpscaling,DISCO}.
We choose additive error as it allows for smaller estimators, and it is similar to the error of common frequency estimation and heavy hitter algorithms~\cite{SpaceSavings,CountMinSketch}. That is,
additive error is unavoidable even if we integrate multiplicative counters into such algorithms. Another argument for additive error is that our estimator size is independent of $N$ while the size of multiplicative error estimators cannot be independent of $N$. 

\subsection{Unit Weight Estimators}
\label{sec:single}

A unit weight estimator supports the {\sc Probabilistic Increment (PIncrement}, or in short PI) and {\sc Query} methods. 
The {\sc PIncrement} method 
adds one to our estimator with a (fixed) probability $p$ which we determine below.  The {\sc Query} method
estimates the number of PIs attempted by returning the value $C/p$ where $C$ is the estimator value. 
%
To determine $p$ we first set $N'=\ceil{2\cdot(1+\epsilon/3)\cdot\epsilon^{-2}\cdot{\ln2\delta^{-1}}}$, and $p=\frac{N'}{N}$.  

Since we know that the maximal query return value is $N$, our estimator only need to count to $N\cdot p =N'$. Intuitively, if we want to increase the estimator above $N'$ it is always due to oversampling.  As a result, we require $\ceil{\log_2(1+N')}\approx 2\log_2\epsilon^{-1} + \log_2\log_2\delta^{-1}+1$ bits. 
Note that the number of bits we require to count until $N$ (estimator value of $N'$) with an additive error of $\epsilon \cdot N$ is independent of $N$. That is, our estimators have an unbounded counting range within the additive error model (note that the  error in the additive model depends on $N$). 
We note that representing $p$ requires $\Omega(\log N)$ bits which implies that our memory consumption still depends on $N$. However, when we move to using arrays of these estimators, since all of the estimators use the same $p$, encoding $p$ introduces a negligible overhead. 

Theorem~\ref{thm:goldbach} shows that our estimation method has the desired property. 
The proof is delayed to Appendix~\ref{app:singleCounterProof}.
\begin{restatable*}[Single Estimator]{thm}{single}
\label{thm:goldbach}
For any number of probabilistic increments $I\le N$, \mbox{we have $\Pr[|C/p-I|> N\epsilon]\le\delta.$}
\end{restatable*}

As an example, Theorem~\ref{thm:goldbach} implies that a $24$-bit estimator can approximate any count up to any pre-specified $N$ within an additive error of $N\epsilon$ for $\epsilon=0.1\%$, and be correct with probability $(1-\delta)$ of $99.95\%$.



\subsection{Weighted Estimators}\label{sec:weighted}
We now consider a weighted estimator where the desired increment can be an arbitrary number (and not just by 1). Such estimators are useful for applications that, for example, rely on the byte volume of flows rather than their packet counts. Further, most existing sketches (e.g., Count Min~\cite{CountMinSketch} and Count Sketch~\cite{CountSketch}) and counter-based algorithms (including Space Saving~\cite{SpaceSavings}, Frequent~\cite{BatchDecrement,frequent4} and RAP~\cite{RAP}) support weighted updates.
\mbox{The recent estimators by~\cite{Infocom2019} support it as well.}

Our weighted estimator supports the {\sc Add}($\mathfrak w$) method, and the {\sc Query} method estimates the sum of all add operations. For example, {\sc PIncrement} is equivalent to {\sc Add}($1$). We generalize $N$ to be the sum of all add operations when discussing weighted measurements.  The notation $N'$ and $p$ are unchanged. 

In the {\sc Add}($\mathfrak w$) method, we break the update into two parts.  Let $w_1 = \floor{\mathfrak w p}$ and $w_2 = \mathfrak w - w_1/p$. We
increase the estimator (deterministically) by $w_1$, and with a probability of $w_2 p$ (notice that $w_2<1/p$ and this is a valid probability), we further increase the estimator by 1. 
In Appendix~\ref{app:weightedUpdatesProof} we prove the correctness of this approach. 

\subsection{Estimator Arrays}
We now discuss how to efficiently implement an estimator array, which is an important building block for sketch algorithms. 
An \emph{estimator array} supports the {\sc PIncrement}$(i)$ and {\sc Query}$(i)$ methods, for $i\in\set{1,\ldots,w}$. Here, $w$ is the number of estimators in the array, also referred to as its \emph{width}.
$N$ is then defined as the overall number of probabilistic increments across all $i$'s and the goal is to estimate the number of {\sc PIncrement}$(i)$'s to within an $N\epsilon$ additive error.

We can further reduce the size of the array since the sum of all
estimators is unlikely to be much larger than $N'$, as an estimator value of $N'$ yields an estimation of $N$. 
Specifically, in Appendix~\ref{app:sumOfCountersBoundProof} we prove that the total number of actual increments to the array is at most $\widetilde{N'}\triangleq N'+\sqrt{3N'\ln\delta_o^{-1}}$ with probability $1-\delta_o$ (the $o$ subscript denotes \emph{oversampling} error probability to distinguish it from the other error sources).

Our goal is to use shorter estimators, and to do so we consider a threshold value $T<N'$, such that each estimator is $\ceil{\log_2 T}$ bits long. \emph{Heavy estimators} are ones which reach the maximal estimator value of $T$, these counters \emph{overflow} to a secondary data structure. Since we keep the sum of all counters bounded by $\widetilde{N'}$, \mbox{there can be at most $\floor{\widetilde{N'}/T}$ heavy counters.}

We store the list of heavy estimators in a hash table where the key is the index of the heavy estimator and the value contains the most significant bits of that estimator. For example, if $N' = 2^{24}-1$, we can have two byte (16 bit) estimators, and extend estimators that require more than 16 bits with another 8 bits. 
In practice, we suggest storing the heavy counters in  a compact hash table such as~\cite{TinyTable,TinyTable2} which adds an additional $\log_2 w + O(1)$ bits per heavy counter or $\floor{\widetilde{N'}/T}(\log_2 w + O(1))$ bits overall.
This means that our total space requirement is $w\ceil{\log_2 T}+\floor{\widetilde{N'}/T}(\log_2 w + O(1))$. 
We minimize this quantity by setting $\delta_o \ll \delta$ and $T\approx \frac{\widetilde{N'}\log_2 w}{w\log_2\parentheses{ \frac{N'\log_2 w}{w}}}$ which gives a total space of $w\cdotpa {\log_2{\frac{N'\log_2 w}{w}}+O(1)}$ bits.~\footnote{For performance, it may be better to set $T=2^{8z}$ for some integer parameter $z$. This allows byte alignment and faster implementation.} That is, we save nearly $\log_2 w$ bits per counter by encoding the heavy ones separately. 
For example, if $w=1024, \epsilon=0.1\%$ and $\delta=99.95\%$, we can set $T=2^{16}$ to encode each counter with two bytes and have at most $253$ heavy counters (even if $\delta_o=2\cdot10^{-15}$), for a total memory of less than $2.5$KB. 
In comparison, allocating 3 bytes for each counter, as in the previous sections, requires $3B\cdot2^{10}=3$KB  (20\% more space).  

\subsection{Dynamically increasing $N$}
\label{sec:dynamicN}
Heretofore, we have assumed that $N$ is known, which allowed us to tune our sampling rate $p$. 
Sometimes $N$ may not be known in advance (e.g., in the case where the measurement length is defined in time and not packets). We propose two algorithms for such a scenario -- {\sc MaxAccuracy} and {\sc MaxSpeed}. Intuitively, {\sc MaxAccuracy} aims for the best accuracy possible given the counter size, while {\sc MaxSpeed} uses the minimal sampling probability to preserve the accuracy guarantee and is therefore faster.

In {\sc MaxAccuracy}, we start with $p=1$, and whenever some counter needs to exceed its maximal value we
independently replace each $C$-valued counter with a generated binomial random variable $\mbox{Bin}(C,1/2)$ and halve the value of $p$. This procedure is called \emph{downsampling} and was first introduced in~\cite{gibbons1998new}.
That is, once \emph{some} counter overflows we decrease the value of \emph{all} counters.
This simulates a process where each {\sc PIncrement} increased the value of the estimator with the current value of $p$. As a result, our accuracy guarantees seamlessly follow for the new estimator, given that $\epsilon,\delta$ are such that $N'$ is smaller than $2^\ell$ for estimators of length $\ell$.
For example, if we are using $\ell=16$-bit counters, then once a counter is incremented for the $(2^{16})$'th time, we halve $p$ and downsample the estimator. 

{\sc MaxSpeed} does not wait for a counter to reach its maximal value, but instead tracks the number of PIs, which we denote by $n$, and uses a sampling probability $\min\set{1,2^{-\floor{\log_2 (n/ N')}}}$.
That is, the first $2N'$ PIs are performed with probability $1$, the next $2N'$ PIs with probability $1/2$, then for $4N'$ PIs it is reduced to $1/4$, etc.
Whenever we halve the sampling probability, we also downsample the counter to maintain the accuracy guarantees. We note that this estimator requires $\approx \log_2(2N')=1+\log_2 N'$ bits, i.e., one additional bit compared to our \mbox{estimator when knowing $N$ in advance.}

The pseudocode for {\sc MaxAccuracy} is given in Algorithm~\ref{alg:maxACC}, and for {\sc MaxSpeed} in Algorithm~\ref{alg:maxSpeed}.  These are generic algorithms that apply to many sketch and cache-based algorithms. Such algorithms vary in the way they implement Line~\ref{linegenaccuracy} in Algorithm~\ref{alg:maxACC}, and Line~\ref{linegenspeed} in Algorithm~\ref{alg:maxSpeed}. The line returns the counters of $x$, which are algorithm dependent. For example, in the CM Sketch~\cite{CountMinSketch} and the CU Sketch~\cite{CUSketch} the set contains a single counter from each array chosen by applying a hash function to $x$. In Space Saving~\cite{SpaceSavings} and RAP~\cite{RAP}, the counter is $x's$ counter if it is monitored, or the minimal counter if it is not monitored. 
Notice that the algorithms may take steps in addition to increasing the counters using our algorithm. For example, Space Saving and RAP may replace the identifier associated with the minimal counter in addition to increasing it.

\begin{figure}[h]
\vspace*{-8mm}
\subfloat[8-bit estimators]
{\includegraphics[width =0.495\columnwidth, height=3.2cm]
{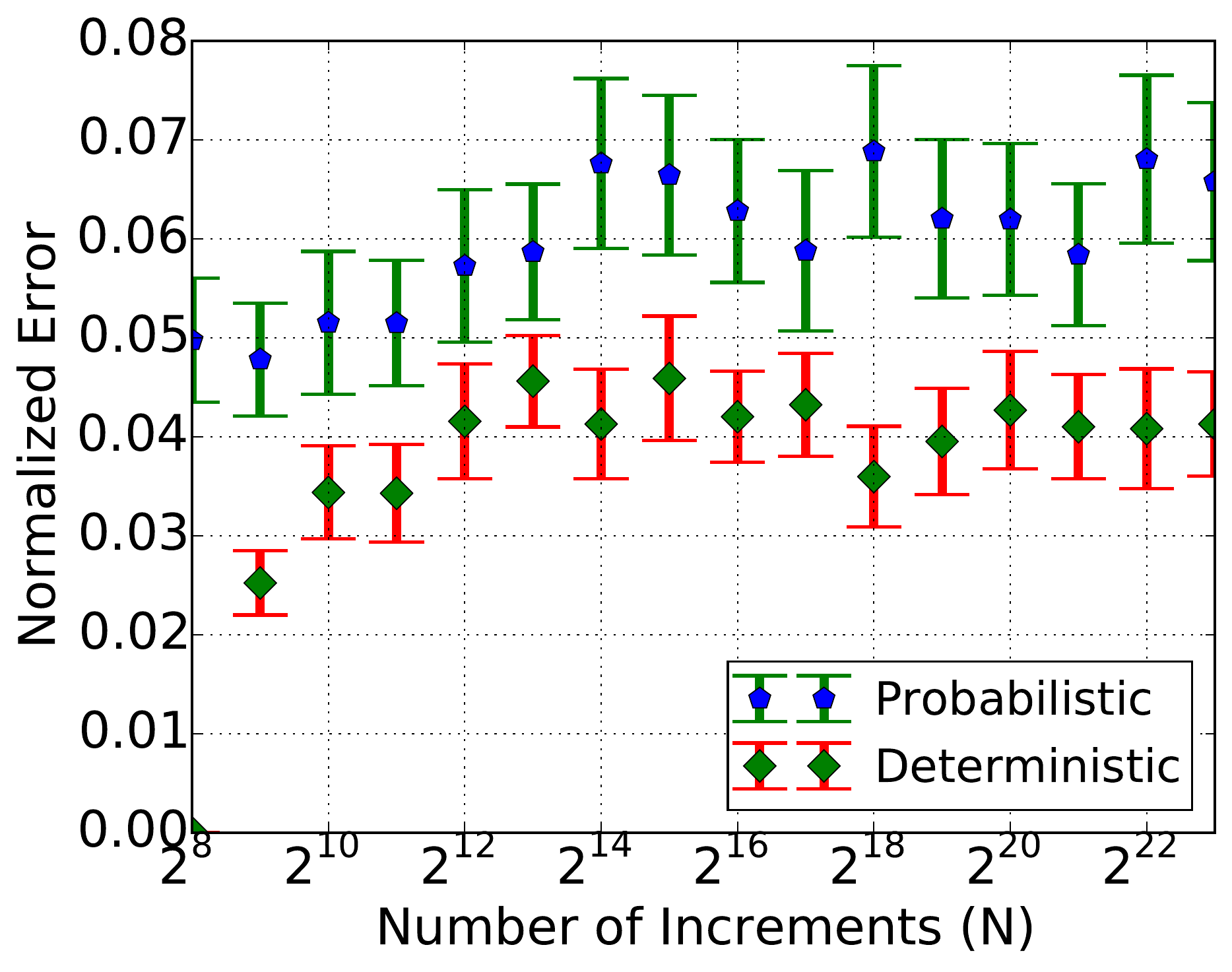}}
\subfloat[16-bit estimators]
{\includegraphics[width =0.495\columnwidth, height=3.2cm]
{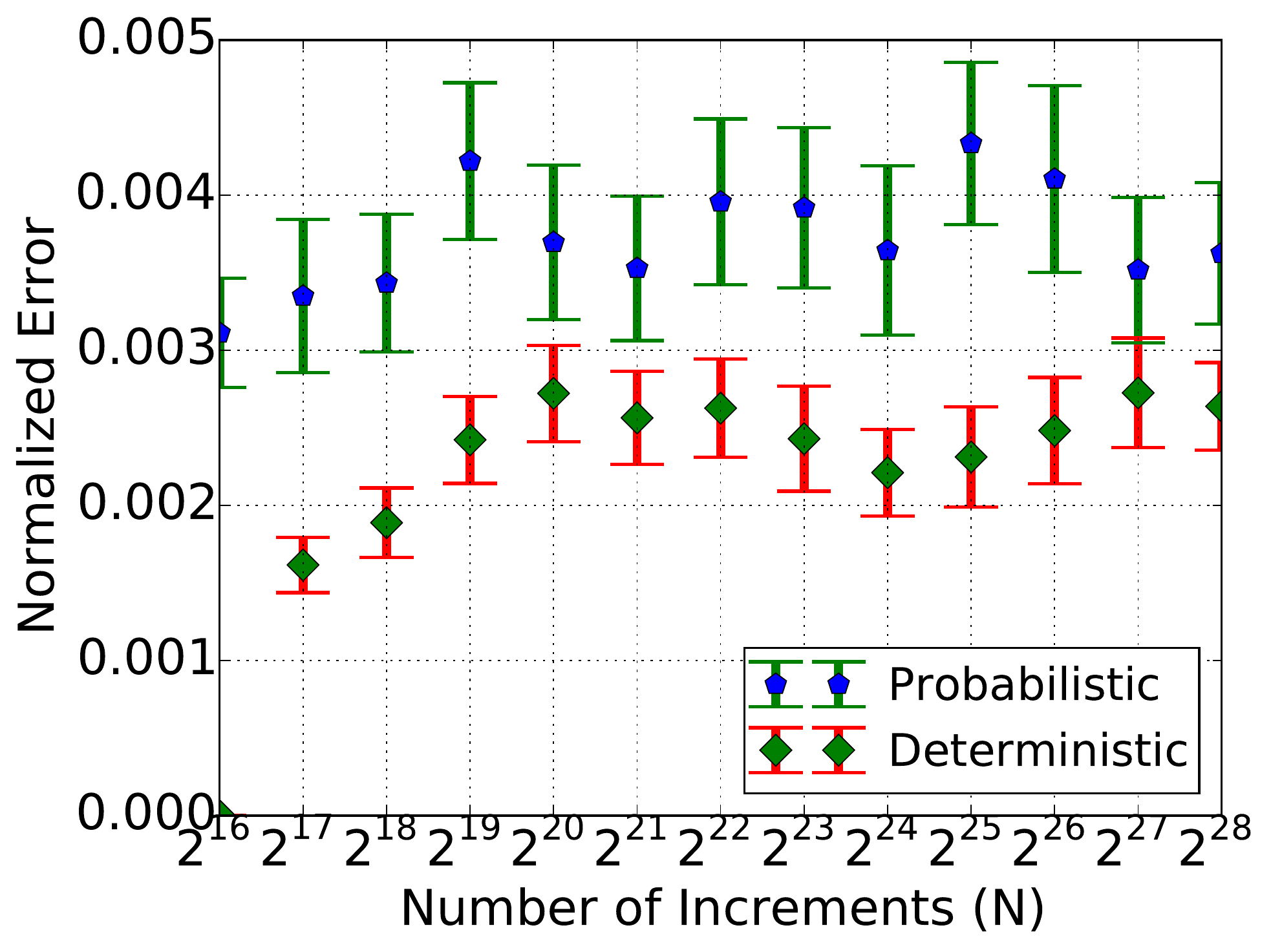}}
\vspace*{-1mm}
\caption{\small \mbox{Comparing probabilistic and deterministic downsampling.\label{fig:downsampling}}}
\vspace*{-3mm}
\end{figure}

\noindent\textbf{Deterministic Downsampling.}\quad{}
We now propose a deterministic method for reducing the estimator values (in both {\sc MaxAccuracy} and {\sc MaxSpeed}). Specifically, when downsampling a $C$-valued estimator, we replace its value with $\floor{C/2}$ instead of $\mbox{Bin}(C,1/2)$.\footnote{One can get slightly more accurate results by randomized rounding up the estimator by $1$ with probability 50\% if $C$ was odd. However, as this improvement is negligible compared with the error of the estimator we \mbox{eschew it for faster implementation.}} The intuition is that this allows us to reduce the variance in the estimation.
We have run experiments to confirm that the accuracy of the deterministic downsampling is superior to that of the probabilistic one. The theoretical accuracy guarantee of the deterministic downsampling is left for future work.
The experiments, whose results are depicted in Figure~\ref{fig:downsampling}, are obtained by running each point $100$ times and reporting its 95\% interval according to Student t-test~\cite{student1908probable}. As shown, the \mbox{deterministic downsampling is indeed more accurate.}

\noindent\textbf{Deamortized Downsampling.}\quad{} Both algorithm variants include a downsampling operation that requires linear time. In some deployments, having a long maintenance operation may cause high latency and even packet drops. To deamortize the downsampling operation and ensure low worst-case update time, we add a \emph{generation} bit to each counter, which specifies if it was downsampled an even number of times. Then, for each packet, we downsample a number of counters that asymptotically equals the amortized update time (e.g., with $2^{18}$ sixteen-bit counters, we can downsample $8$ counters in each update). Importantly, if a counter that has not been downsampled yet overflows, we immediately downsample it and switch its generation bit, \mbox{to identify it once the maintenance operation reaches it.}

\subsection{Optimizing the Update Speed}
While our proposed estimator saves space, we designed it in a manner that can also reduce the update time. The key aspect of our approach is that the probability for updating an estimator \emph{does not} depend on its current value. In comparison, the update probability in all the estimator techniques surveyed in this work~\cite{ICE-Buckets,CASE,CEDAR,ANLSUpscaling,DISCO,SAC,Infocom2019} depends \mbox{on the current estimator value.}

Specifically, we can decide if an estimator is updated prior to calculating the sketch hash functions, and without reading any data structure. When $N$ is large enough, most packets require no additional work as they do not update any estimator. 
Further, we can use Geometric Sampling~\cite{Nitro} to determine how many packets to skip before an estimator is updated. If each packet is sampled with probability $p$, then the number of packets until the next sample is distributed geometrically with mean $p^{-1}$. Geometric Sampling simply generates \mbox{a single} variable $G\sim \mbox{Geo}(p)$ (i.e., $\Pr[G=x]=p(1-p)^{x-1}$) by using the Inverse Transform Sampling method.
The method sets $G=\ln U / \ln (1-p)$ for a uniform random variable $U\sim\mbox{Uniform}[0,1]$; it requires a single uniform variate and a few floating-point operations. The variable $G$ is shared across \emph{all} estimators and thus does not impose a significant memory overhead (e.g., it can be implemented as a 64-bit integer). 
While a similar approach for acceleration appears in NitroSketch~\cite{Nitro}, it does not allow for shorter counters as they add $p^{-1}$ to the sampled counters and vary $p$ over time. 

For sketches that associate each flow with $d$ estimators, such as the Count Min Sketch and Conservative update,
the geometric sampling  only requires $d\cdot N'$ operations per $N$ packets, which gives an amortized complexity of $1+\frac{d\cdot N'}{N}=O\parentheses{1+\frac{\epsilon^{-2}\log^2\delta^{-1}}{N}}$. That is, we have a constant update time for streams in which $N=\Omega(\epsilon^{-2}\log^2\delta^{-1})$.

While cache-based algorithms such as Space-Saving and Frequent have data structures that allow constant-time updates~\cite{CormodeCode}, they may require seven pointers per entry. 
Alternative approaches include a heap-implementation~\cite{CormodeCode} that, while being space-efficient, requires a logarithmic update time.
Our approach allows using a heap while keeping the amortized update complexity constant (in streams in which $N=\Omega(\epsilon^{-2}\log\epsilon^{-1}\log\delta^{-1})$).

\begin{algorithm}[ht]
\small
\caption {{\sc MaxAccuracy} Algorithm with $n$-bits counter\label{alg:maxACC}}
\begin{algorithmic}[1]
\Statex Initialization: $p\gets 1$
\Procedure{Add}{$\langle x,w\rangle$}\Comment{A $w$-sized packet from flow $x$}
    \State $w_1\gets \floor{w\cdot p}$
    ,\quad{}
    $w_2\gets w-w_1$
    \State \label{linegenaccuracy}$\mathcal C\gets \mbox{$x$'s counters}$ \Comment{Algorithm dependent}
    \While {$\max\set{\mathcal C} \ge 2^n-w_1$}\Comment{Overflow event}\label{line:maxACCOverflow1}
        \State Divide \emph{all} counters by two \Comment{Not only $x$'s counters}
        \State $p\gets p/2$
        \State $w_1\gets \floor{w\cdot p}$
        ,\quad{}
        $w_2\gets w-w_1$\label{line:endMaxACCOverflow1}
    \EndWhile
    \For {${C\in \mathcal C}$}
        \State $C\gets C + w_1$
        \If {$U[0,1] \le w_2\cdot p$}\Comment{With probability $w_2\cdot p$}
            \If {$C = 2^n-1$}\Comment{Overflow event}
                \State Divide all counters by two
                \State $p\gets p/2$
                \State $w_1\gets \floor{w\cdot p}$
        ,\quad{}
        $w_2\gets w-w_1$
            \EndIf
            \State $C\gets C+1$
        \EndIf
    \EndFor
\EndProcedure
\Procedure{Query}{x}\Comment{Estimate flow $x$'s size}
    \State $q \gets$ Algorithm's estimate\Comment{Algorithm dependent}
    \State\Return $q/p$
\EndProcedure
\end{algorithmic}
\end{algorithm}

\begin{algorithm}[ht]
\small
\caption {{\sc MaxSpeed} Algorithm with $n$-bits counter\label{alg:maxSpeed}}
\begin{algorithmic}[1]
\Statex Initialization: $p\gets 1, N\gets 0, g\gets 1$
\Statex \qquad{}\qquad{}\qquad{} $N'\gets\ceil{2\cdot(1+\epsilon/3)\cdot\epsilon^{-2}\cdot{\ln2\delta^{-1}}}$
\Procedure{Add}{$\langle x,w\rangle$}\Comment{A $w$-sized packet from flow $x$}
    \State $n\gets n + w$
    \State $p_{\mbox{new}}\gets 2^{-\floor{\log_2 n/N'}}$
    \If {$p_{\mbox{new}} < p$}
        \State $D=\log_2 (p/p_{\mbox{new}})$
        \State Divide all counters by $2^D$
        \State $p\gets p_{\mbox{new}}$
        \State $G\gets \mathit{Geo}(p)$ \Comment{Geometric random variable}
    \EndIf
    \State $w_1\gets \floor{w\cdot p}$
    ,\quad{}
    $w_2\gets w-w_1$
    \State \label{linegenspeed}$\mathcal C\gets \mbox{$x$'s counters}$ \Comment{Algorithm dependent}
    \For {${C\in \mathcal C}$}
        \State $C\gets C + w_1$
        \While {$G \le w_2$}\Comment{Simulate $w_2$ coin flips}
            \State $C\gets C+1$
            \State $w_2\gets w_2 - G$
            \State $G\gets \mathit{Geo}(p)$ \Comment{Geometric random variable}
        \EndWhile
        \State $G\gets G - w_2$
    \EndFor
\EndProcedure
\Procedure{Query}{x}\Comment{Estimate flow $x$'s size}
    \State $q \gets$ Algorithm's estimate\Comment{Algorithm dependent}
    \State\Return $q/p$
\EndProcedure
\end{algorithmic}
\end{algorithm}
\section{Integrating Estimator Arrays with Sketches}\label{sec:counterArrays}
Sketch data structures utilize several independent counter arrays. 
Intuitively, each array provides an estimation which is (roughly) accurate with a constant probability, and additional arrays amplify the success probability.
For example, the Count Min Sketch (CMS)~\cite{CountMinSketch} employs $d=O(\log\delta^{-1})$ arrays $A_1,\ldots,A_d$ of $w=O(\epsilon^{-1})$ counters each. Whenever an element $x$ arrives, it uses $d$ uncorrelated pairwise-independent hash functions $h_1,\ldots,h_d$ that map the input to the range $[0,w)$, and for each $j=1,\ldots,d$ it increments the counter $A_j[h_j(x)]$ of the $j$'th array.
When receiving a query for the multiplicity of $x$, we take the minimum over all $j$ of $A_j[h_j(x)]$.  Clearly, CMS can be implemented using our estimator array algorithm above, replacing increment operations with the probabilistic increment operations. For example, with $d=5$ arrays of $w=2^{10}$ counters each, we require about $12.5$KB for the entire encoding.

The sketch itself also has an error that is caused by collisions of different items that increment the same counter. For CMS, it guarantees that the error will be bounded by $N\epsilon_A$ with probability $1-\delta_A$, for $\epsilon_A = e/w$ and $\delta_A=e^{-d}$. Combining the error from the sketch with that of the counter arrays, we have an error of at most $\epsilon+\epsilon_A$ with probability at least $1-d\cdot \delta-\delta_A$. For example, if $d=5$ and $w=2^{10}$ then replacing the CMS's counters (assuming they are 32-bits each) with our estimators reduces the space from $20$KB to $12.5$KB while increasing the error from 0.271\% to 0.371\% and the error probability from 0.67\% to 0.97\%. We note that a CMS configured for a 0.371\% error except with probability 0.97\% would still require more space ($>13.5$KB) than our solution (while also being considerably slower). 

\section{Cache-based Counter Algorithms}
\label{sec:CBA}
Sketches are a popular design choice for hardware as they are easy to implement in hardware.
In software, however, one can generally get a better accuracy to space tradeoff by using cache-based counter algorithms~\cite{SpaceSavingIsTheBest2010,SpaceSavingIsTheBest2011}. Specifically, algorithms like Space Saving~\cite{SpaceSavings}, Misra-Gries~\cite{misra1982finding}, and Frequent~\cite{BatchDecrement,frequent4} use $O(\epsilon^{-1})$ counters (as opposed to $O(\epsilon^{-1}\log\delta^{-1})$ in sketches such as Count Min). 


In this section, we consider compact cache-based algorithms that can benefit from utilizing estimators, rather than full-sized counters. To obtain maximal benefits, we concurrently aim to minimize the overhead from the flow identifiers. For example, flows are typically defined by five-tuples that are 13 bytes long, whereas counters are typically 4 to 8 bytes long.  In such a setting, reducing a 4-byte counter to a 2-byte estimator offers only marginal space improvements.
We therefore propose replacing the identifiers with \emph{fingerprints}, i.e., short pseudo-random bitstrings generated as hashes of the identifiers. 
Fingerprints were proposed before (e.g., see~\cite{HeavyHitters}) to compress identifiers; however, the following analysis, which asks for the shortest size at which an element experiences additive error at most $N\epsilon$ appears to be new. 
In particular, it allows us to use shorter fingerprints compared to previous analyses.
If the stream contains $D\le N$ distinct items, then fingerprints of size $O(\log D)$ suffice to ensure that no two items have a fingerprint collision (with suitably high probability) and thus the accuracy is essentially unaffected by this compression. However, while fingerprints may be smaller than the $13$ bytes required for encoding five-tuples, they may still be significantly larger than the estimator.
We can do better by not requiring no collisions, and instead finding the minimal fingerprint length ($L$) that allows an error of at most $N\epsilon_f$ with probability $1-\delta_f$.  We show that $L\approx \log\epsilon_f^{-1}\delta_f^{-1}$ suffices, implying that the fingerprint length can be \mbox{of the same order as our estimators.}

We use a weighted variant of the Chernoff bound which states 
that for independent random variables $Y_1,\ldots,Y_n$ with values in the interval $[0,z]$ for some $z>0$, the sum $Y=\sum_{i=1}^n Y_i$ satisfies for all $t>0$, 
$
\Pr[Y > t]\le (e\cdot \mathbb E[Y]/t)^{t/z}.
$

Given a parameter $\alpha\in [0,1]$, we split the items into large and small ones. Let $\mathcal L$  denote the set of items whose size is at least $(\alpha\cdot N\epsilon_f)$, and let $\mathcal S$ denote the remaining. 
Further, let $S_{\mathcal L}$ denote the total size of the large items and let $S_{\mathcal S}$ denote the total size of the small ones. We have that $S_{\mathcal L}+S_{\mathcal S}\le N$.
We want to set the fingerprint size $L$ such that with probability $1-\delta_f$ \emph{none} of the large items collide with $x$ and the sum of sizes for the small colliding items is at most $N\epsilon_f$. 
Using the union bound, and the fact that $|\mathcal L|\le \alpha^{-1}\cdot \frac{S_{\mathcal L}}{N\epsilon_f}$, we have that the probability for a collision with a large item is at most $\alpha^{-1}\cdot\frac{2^{-L}S_{\mathcal L}}{N\epsilon_f}$.
For each small item $i$ with size $f_i\le \alpha\cdot N\epsilon_f$, we define the random variable $Y_i$ to take the value $f_i$ if $i$ has the same fingerprint as $x$ and $0$ otherwise. The total volume that collides with $x$ is then $Y=\sum_{i\in\mathcal S}Y_i$ (i.e., $\mathbb E[Y]=2^{-L}S_{\mathcal S}$). Since each $Y_i$ is bounded by $z=(\alpha\cdot N\epsilon_f)$, we use the Chernoff bound with $t=N\epsilon_f$ to conclude that
{\small \vspace*{-2mm}$$
\Pr[Y > N\epsilon_f]\le (e\cdot \mathbb E[Y]/t)^{t/z}
\le \parentheses{\frac{e\cdot 2^{-L}\cdot S_{\mathcal S}}{N\epsilon_f}}^{\alpha^{-1}}
.
$$\vspace*{-2mm}}

Therefore, the overall chance of failure is at most 
{\small
\begin{align}
\vspace*{-2mm}
\label{eq:ssErrorProb}
\alpha^{-1}\cdot\frac{2^{-L}S_{\mathcal L}}{N\epsilon_f} +  \parentheses{{\frac{e\cdot 2^{-L}\cdot S_{\mathcal S}}{N\epsilon_f}}}^{\alpha^{-1}}.
\vspace*{-2mm}
\end{align}
}

To account for all possible splits of $N$ packets into large and small flows and guarantee that~\eqref{eq:ssErrorProb} is at most $\delta_f$, we choose 
$$L=\ceil{\max\set{\log_2(\alpha^{-1}\cdot\epsilon_f^{-1}\delta_f^{-1}),\log_2(e\cdot\epsilon_f^{-1}\delta_f^{-\alpha})}}$$ to conclude that with probability $1-\delta_f$ at most $N\epsilon_f$ packets collide with the fingerprint of $x$.
For example, by setting $\alpha=10/11$, we find that two byte identifiers yield $\epsilon_f,\delta_f\approx 0.5\%$, three bytes yield an error lower than $\epsilon_f,\delta_f=0.03\%$, and $32$-bit identifiers yield $\epsilon_f,\delta_f < 0.002\%$.

Space Saving, Misra Gries, and Frequent are all deterministic and have an additive error of $N\epsilon_A$, where $\epsilon_A=1/w$ and $w$ is again the width. Therefore, combining them with our estimators (with an $\epsilon,\delta$ guarantee) yields an overall error of $N\cdot(\epsilon+\epsilon_A+\epsilon_f)$ 
with probability at least $1-\delta-\delta_f$.

For brevity, we next provide two numerical examples with $w{=}2^{10}, \epsilon_f{=}0.03\%$ and $\delta_f{=}0.03\%$.

\noindent\textbf{Example 1.} Consider $\epsilon{=}0.1\%, \delta{=}0.05\%$ and $T=2^{16}$; we get an error lower than $0.23\%$ with probability at least $99.92\%$, while compressing the identifiers into three bytes and replacing the counters with two-byte estimators. 
That is, our example requires 5-bytes per entry, compared with $13+4=17$  bytes in the original.
We also have at most $253$ large counters (see Section~\ref{sec:counterArrays}), for an overall memory of $5.5$KB. In contrast, for a $0.23\%$ error guarantee, these algorithms would need $435$ \mbox{counters, requiring more space.}


\noindent\textbf{Example 2.} Consider $\epsilon=2^{-13},\delta=2^{-16}$ and $T=2^{24}$. That is, we require 24-bit estimators and have at most $32$ large estimators. This configuration has a total error of at most $0.14\%$ with probability $\approx 99.97\%$ and requires 6.2KB. In comparison, the uncompressed variants require nearly $14$KB of space for the same guarantees. 
\GI{MM please review}

\section{Evaluation}\label{sec:eval}
We evaluate our algorithms on two real packet traces: the first 98M packets of (1) the CAIDA equinix-newyork 2018 (NY18)~\cite{CAIDA2018} and (2) the CAIDA equinix-newyork 2016 (CH16)~\cite{CAIDA2016} backbone traces. 
We picked these traces as they are somewhat different: CH16 contains 2.5M flows while NY18 exhibits a heavier tail and has nearly 6.5M flows. 
We implement our algorithms in C++ and compare them with the, state of the art, SAC estimators~\cite{Infocom2019} whose code we obtained from the authors. The Baseline code for Space Saving was taken from~\cite{CormodeCode} and we extended it to implement the RAP and dWay-RAP algorithms. For a fair comparison, all algorithms employ the same hash function (BobHash).
The default setting for our algorithm is {\sc MaxAccuracy} and we evaluate the difference from {\sc MaxSpeed} in Section~\ref{sec:maxspeedeval}. 
We ran the evaluation on a PC with an Intel Core i7-7700 CPU @3.60GHz and 16GB DDR3 2133MHz RAM. Finally, we refer to a PI as increment, to be \mbox{consistent across all algorithms.}

\begin{figure}[t]
    \centering
    \subfloat[8-bit estimators, Error]
    {\label{2a} \includegraphics[width =0.25\textwidth]
    {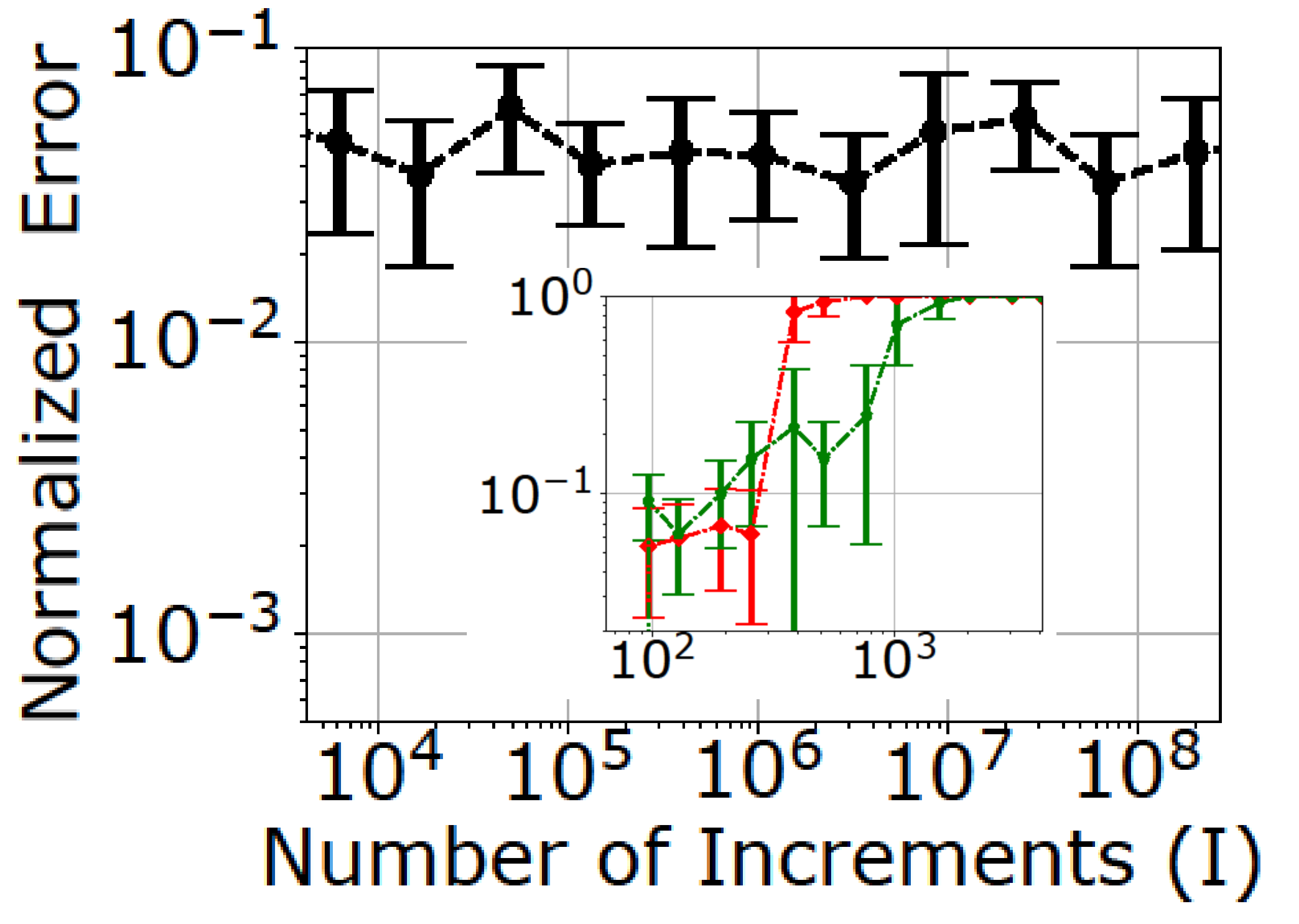}}
    \subfloat[8-bit estimators, Speed]
    {\label{2b}\includegraphics[width =0.25\textwidth]
    {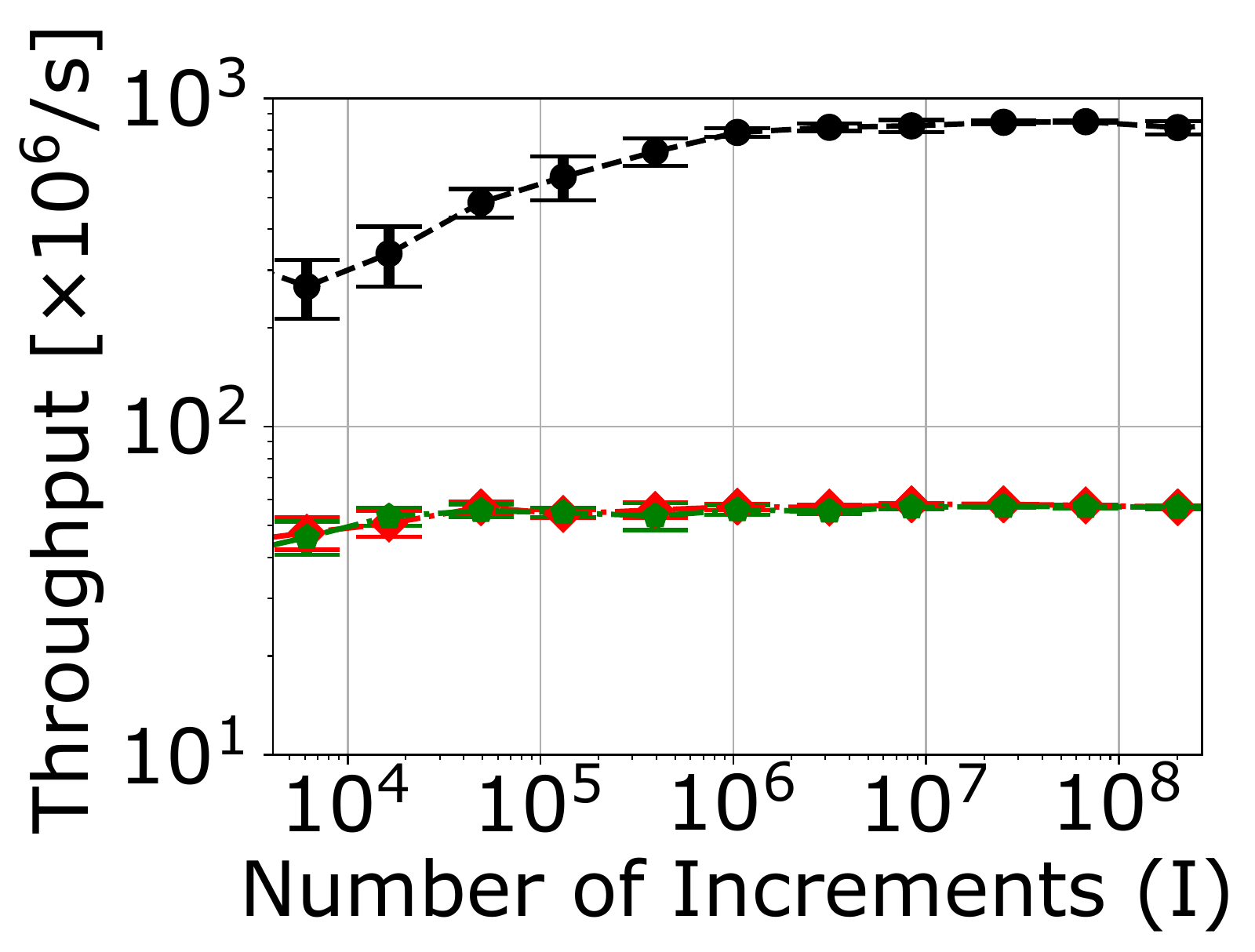}}\\
    \hspace*{1mm}
   {\includegraphics[width =1.0\columnwidth]
    {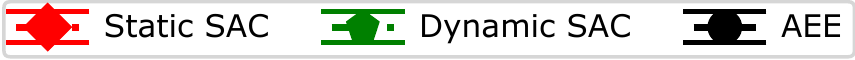}\vspace*{-4mm}}
    \\
    \subfloat[16-bit estimators, Error]
    {\label{2c}\includegraphics[width =0.25\textwidth]
    {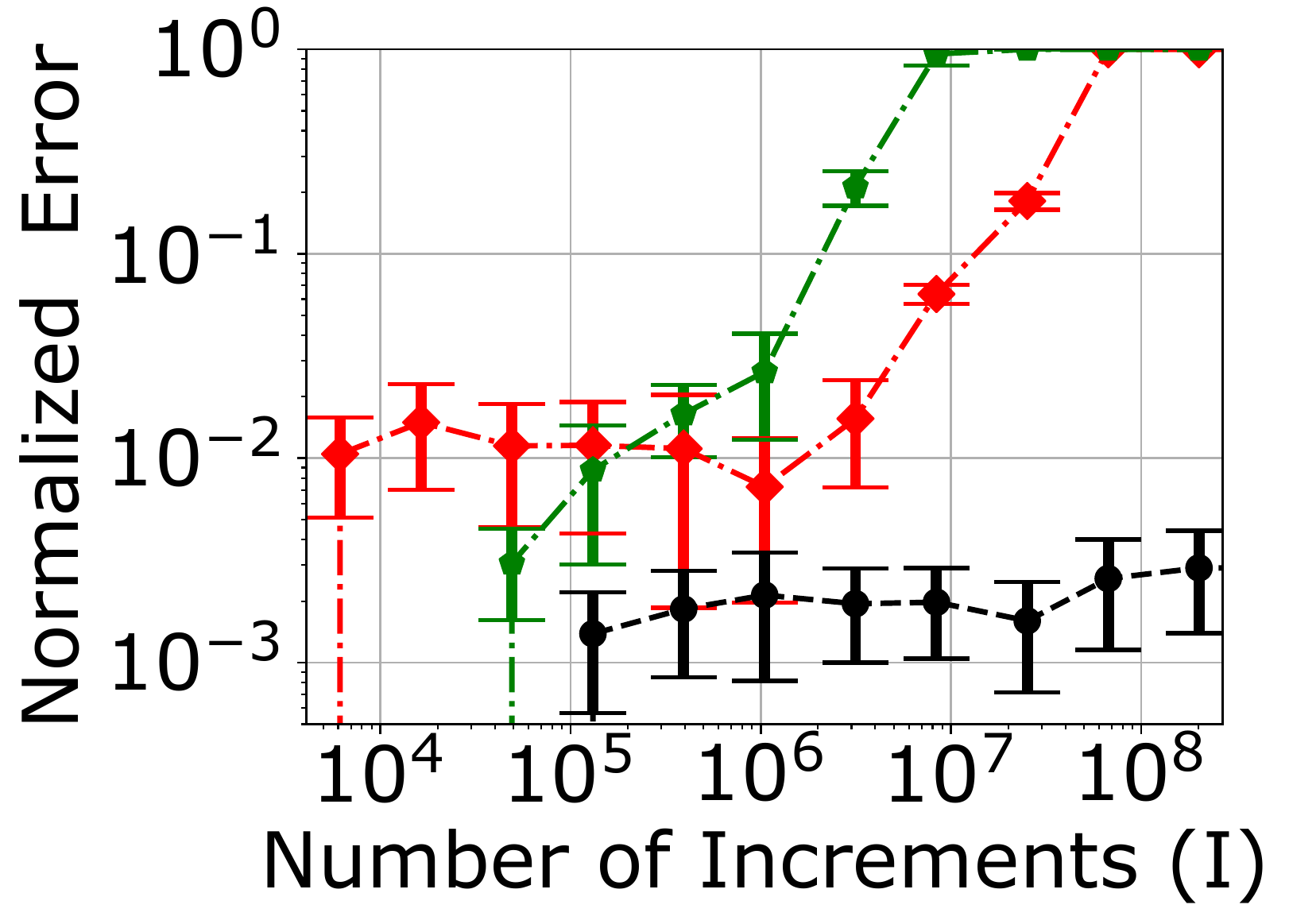}}
    \subfloat[16-bit estimators, Speed]
    {\label{2d}\includegraphics[width =0.25\textwidth]
    {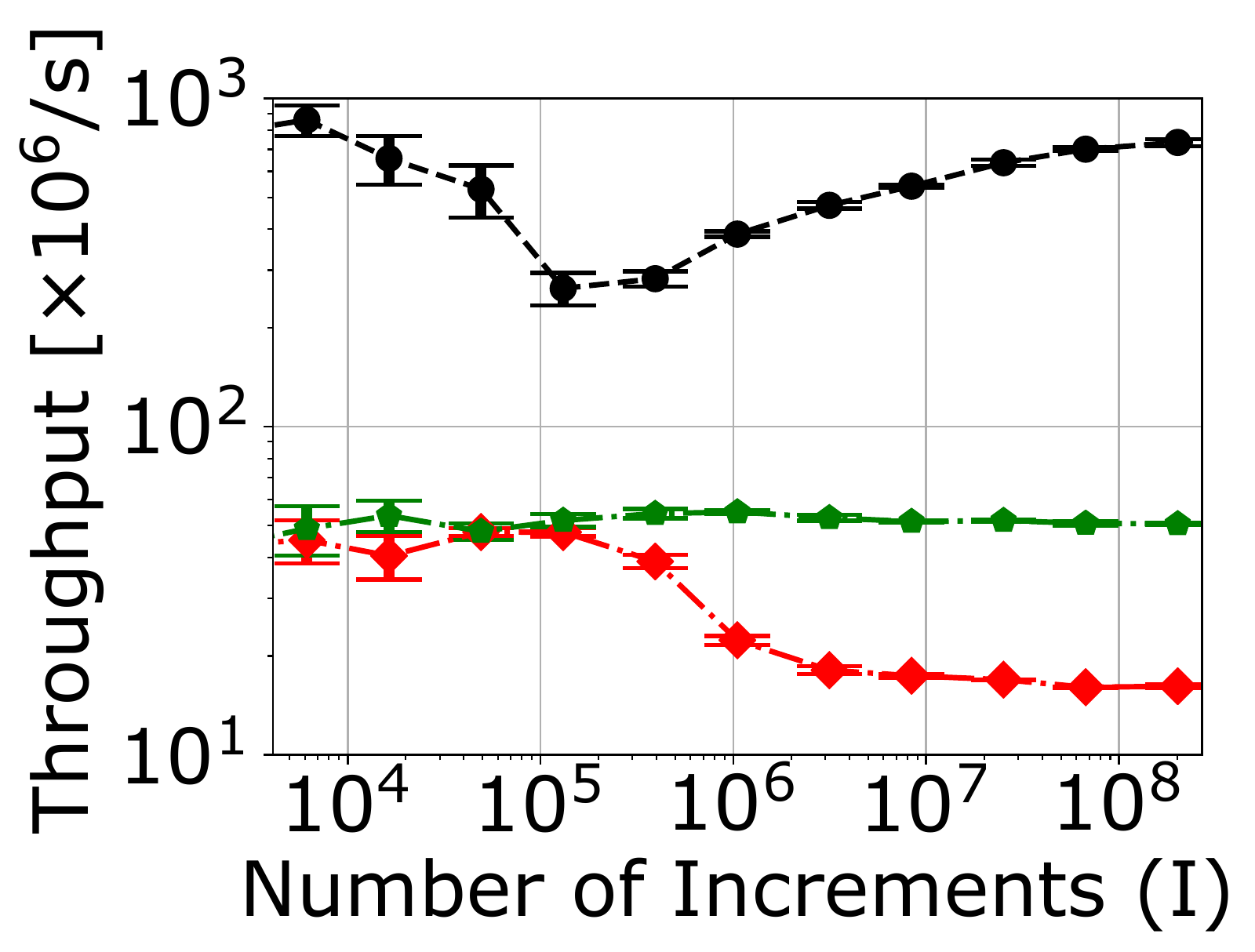}}\\
    \vspace*{-2mm}
    \caption{\small A comparison of the speed and accuracy of single~estimators.}\label{fig:singleCounter} 
    \vspace*{-2mm}
\end{figure}

\begin{figure*}[t]
    \centering
    \hspace*{-3mm}
    \subfloat[CM Sketch, Error, NY18]
    {\label{3a}\includegraphics[width =0.25\textwidth]
    {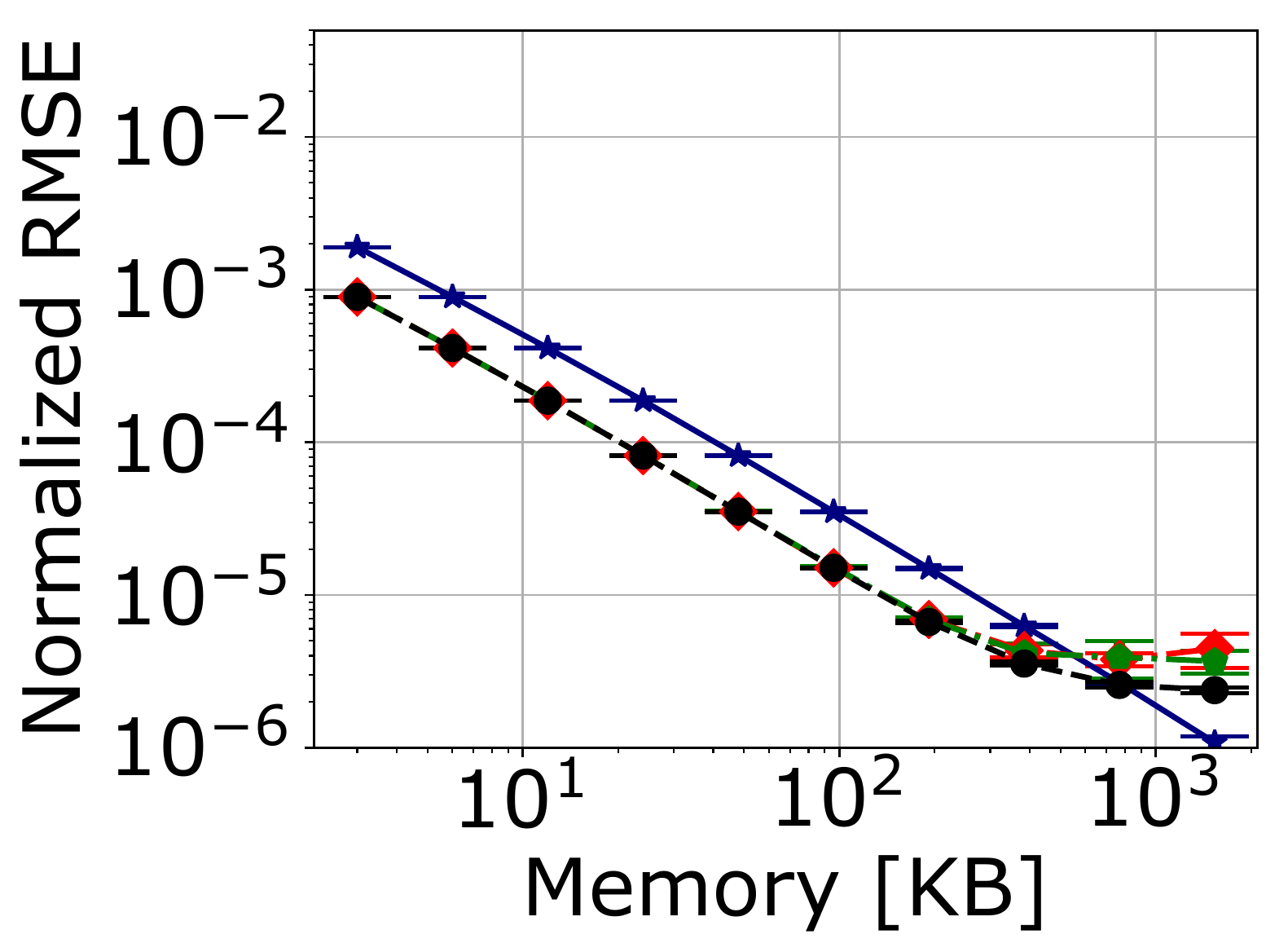}}
    \subfloat[CM Sketch, Error, CH16]
    {\label{3b}\includegraphics[width =0.25\textwidth]
    {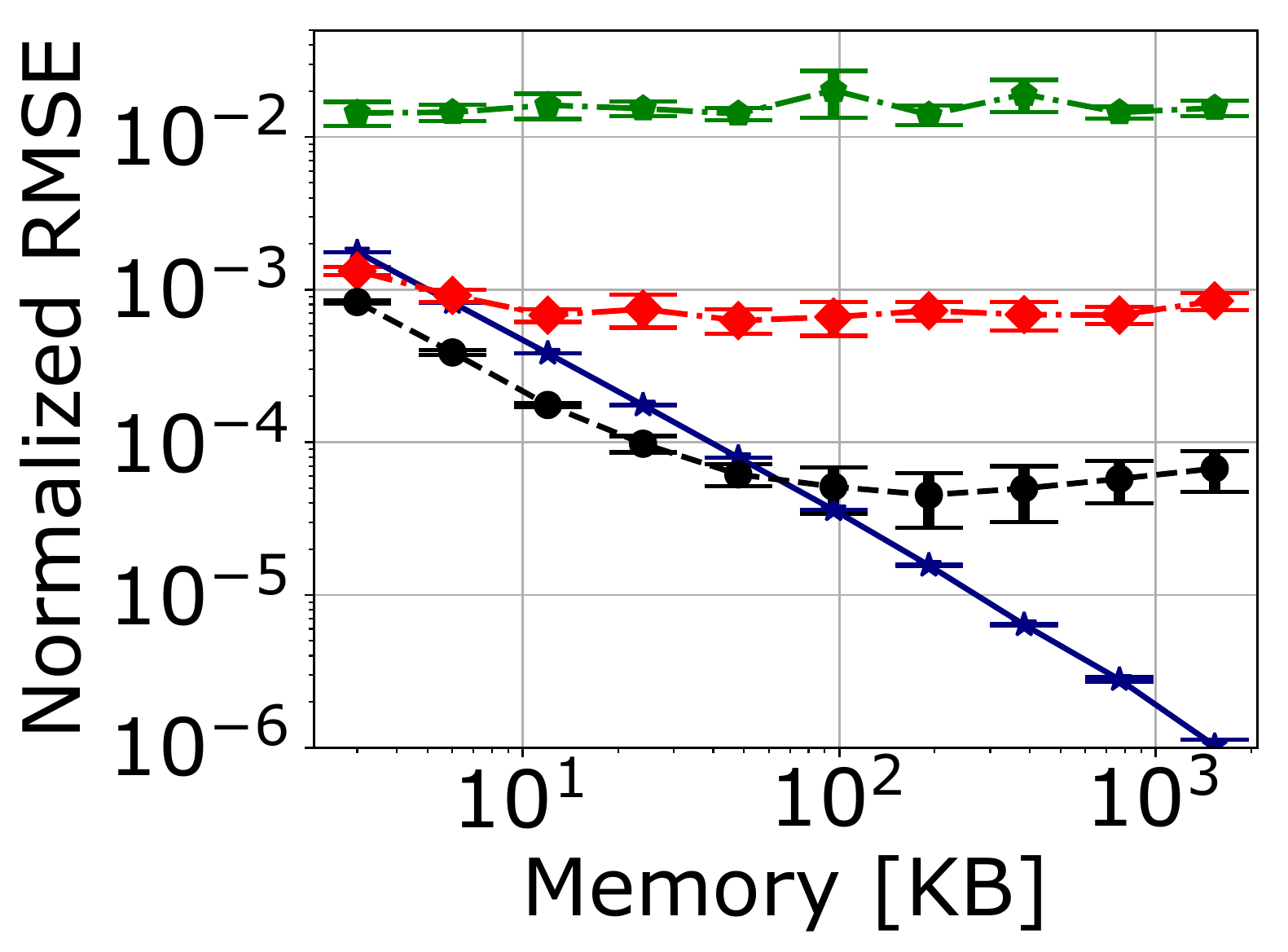}}
    \subfloat[CM Sketch, Speed, NY18]
    {\label{3c}\includegraphics[width =0.25\textwidth]
    {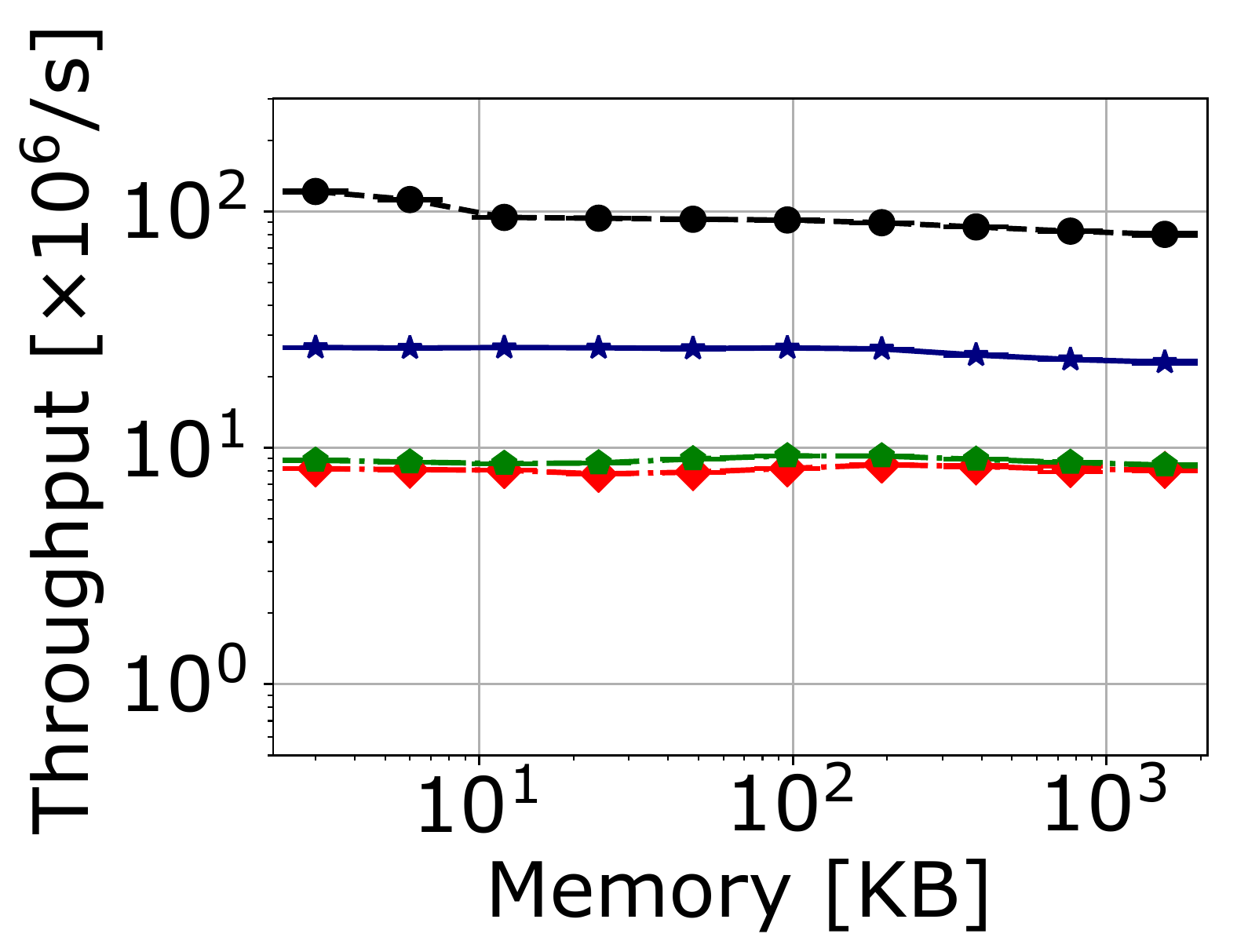}}
    \subfloat[CM Sketch, Speed, CH16]
    {\label{3d}\includegraphics[width =0.25\textwidth]
    {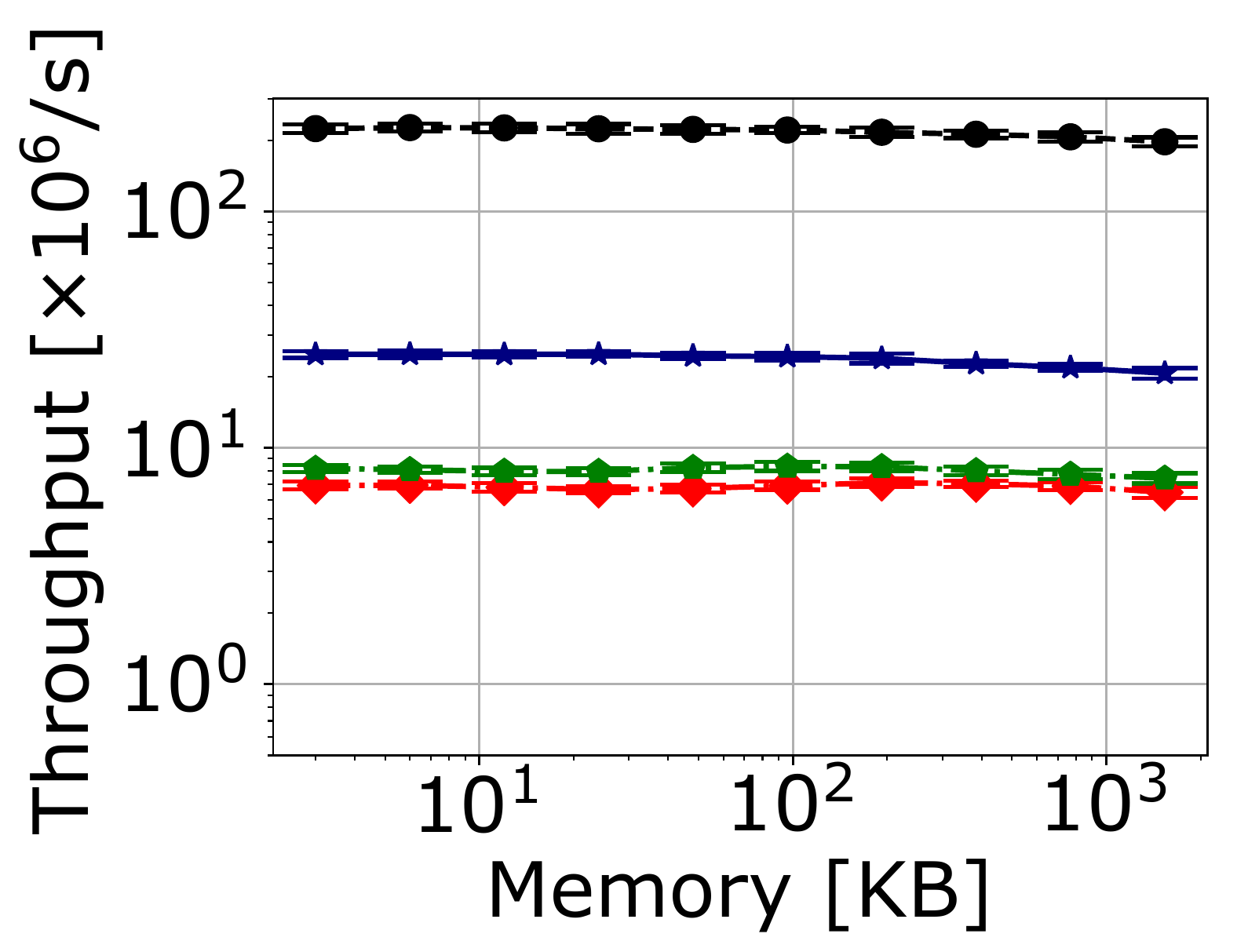}}    \\ 
    {\includegraphics[width =1.2\columnwidth]
    {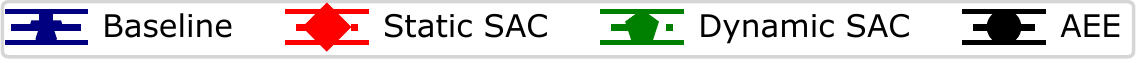}\vspace*{-2mm}}\\
    \hspace*{-3mm}
    \subfloat[CU Sketch, Error, NY18]
    {\label{3e}\includegraphics[width =0.25\textwidth]
    {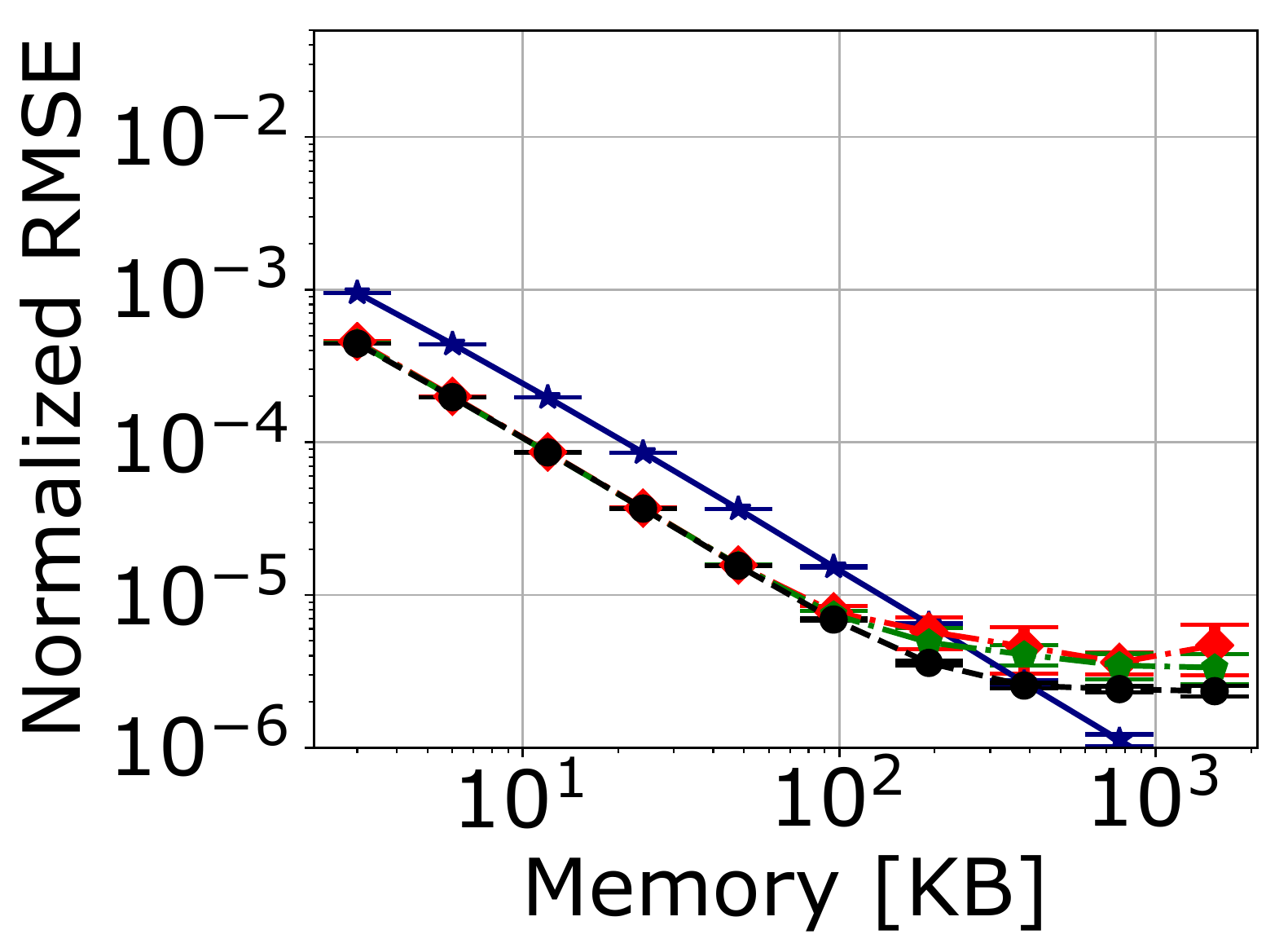}}
    \subfloat[CU Sketch, Error, CH16]
    {\label{3f}\includegraphics[width =0.25\textwidth]
    {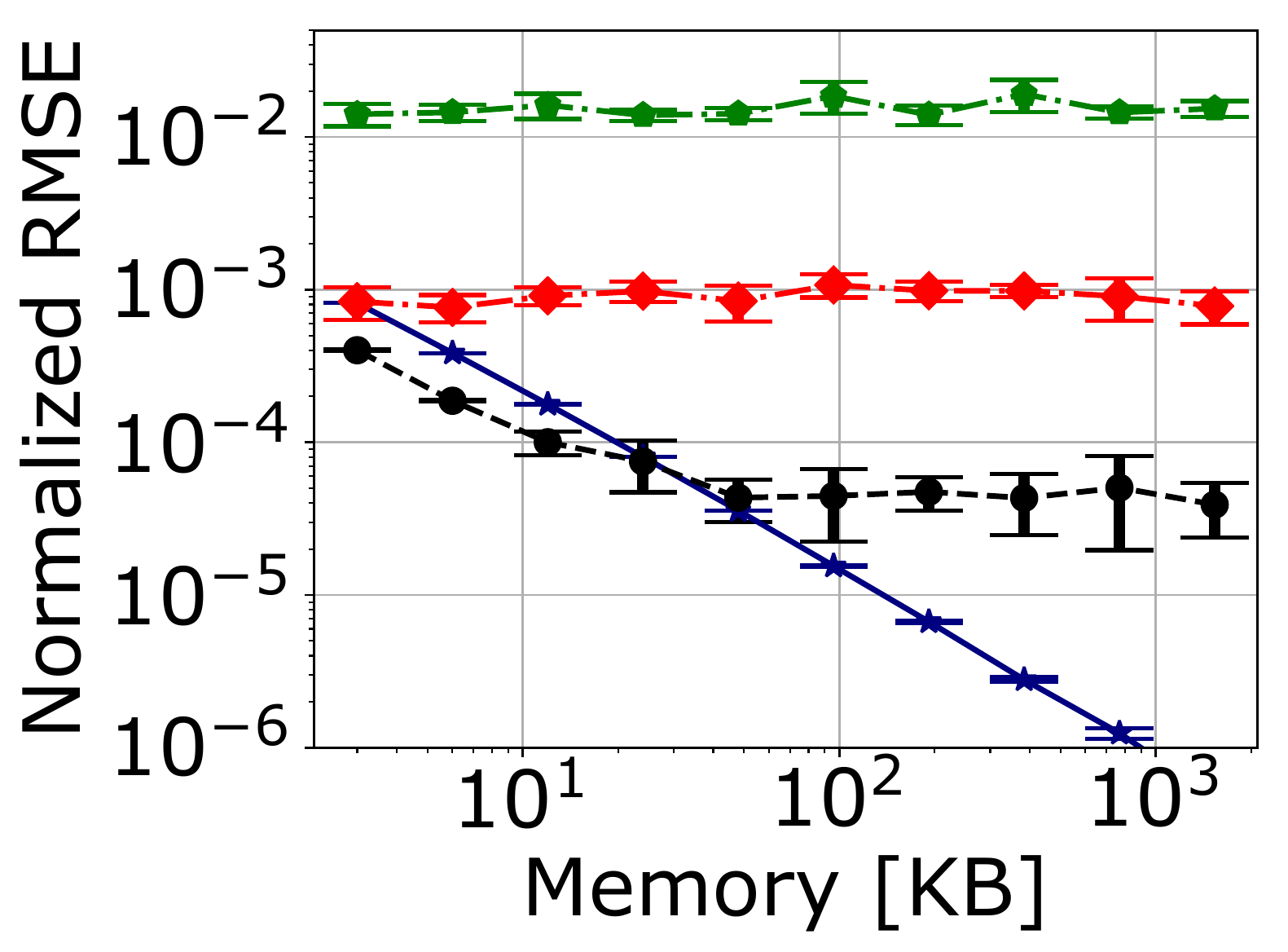}}
    \subfloat[CU Sketch, Speed, NY18]
    {\label{3g}\includegraphics[width =0.25\textwidth]
    {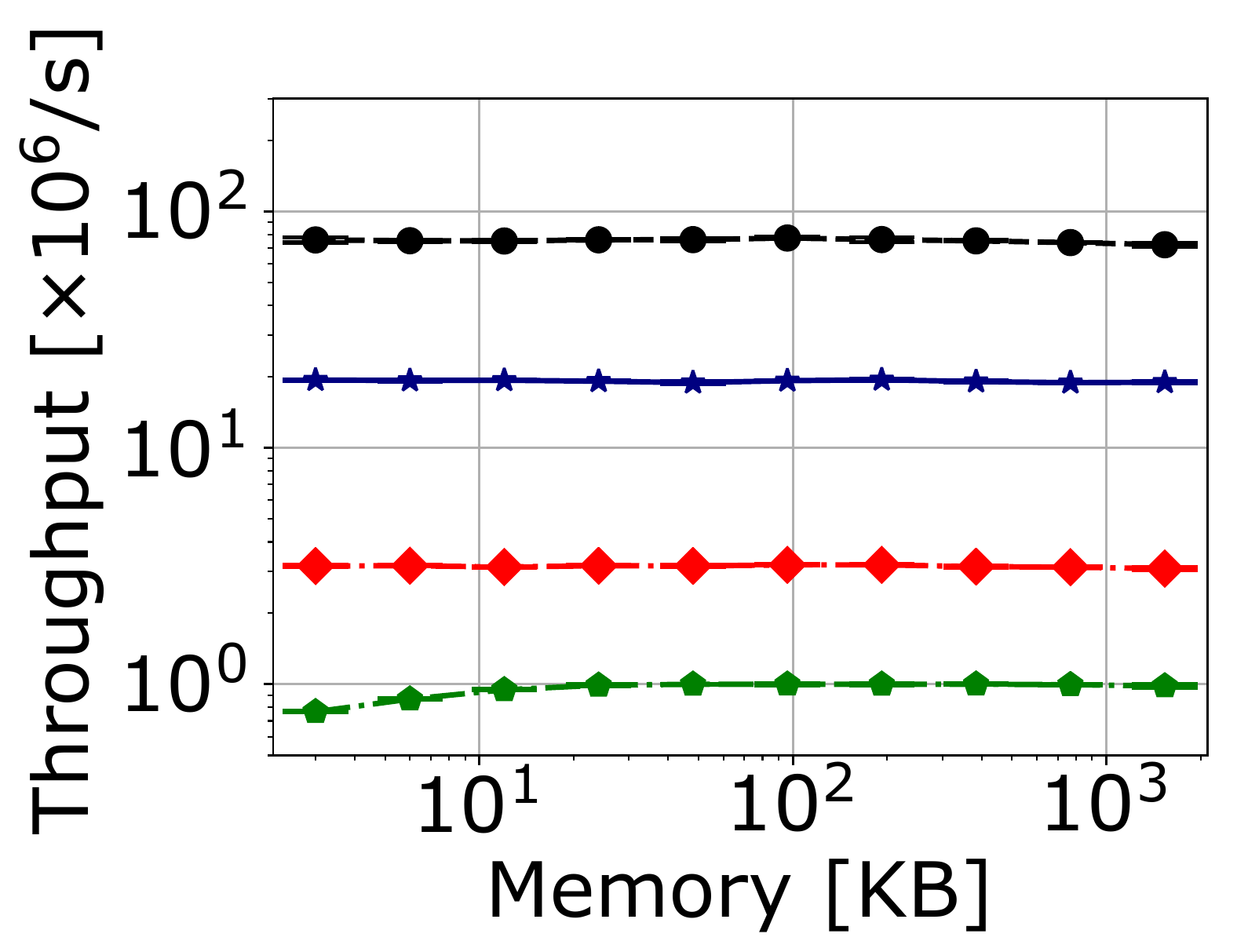}}
    \subfloat[CU Sketch, Speed, CH16]
    {\label{3h}\includegraphics[width =0.25\textwidth]
    {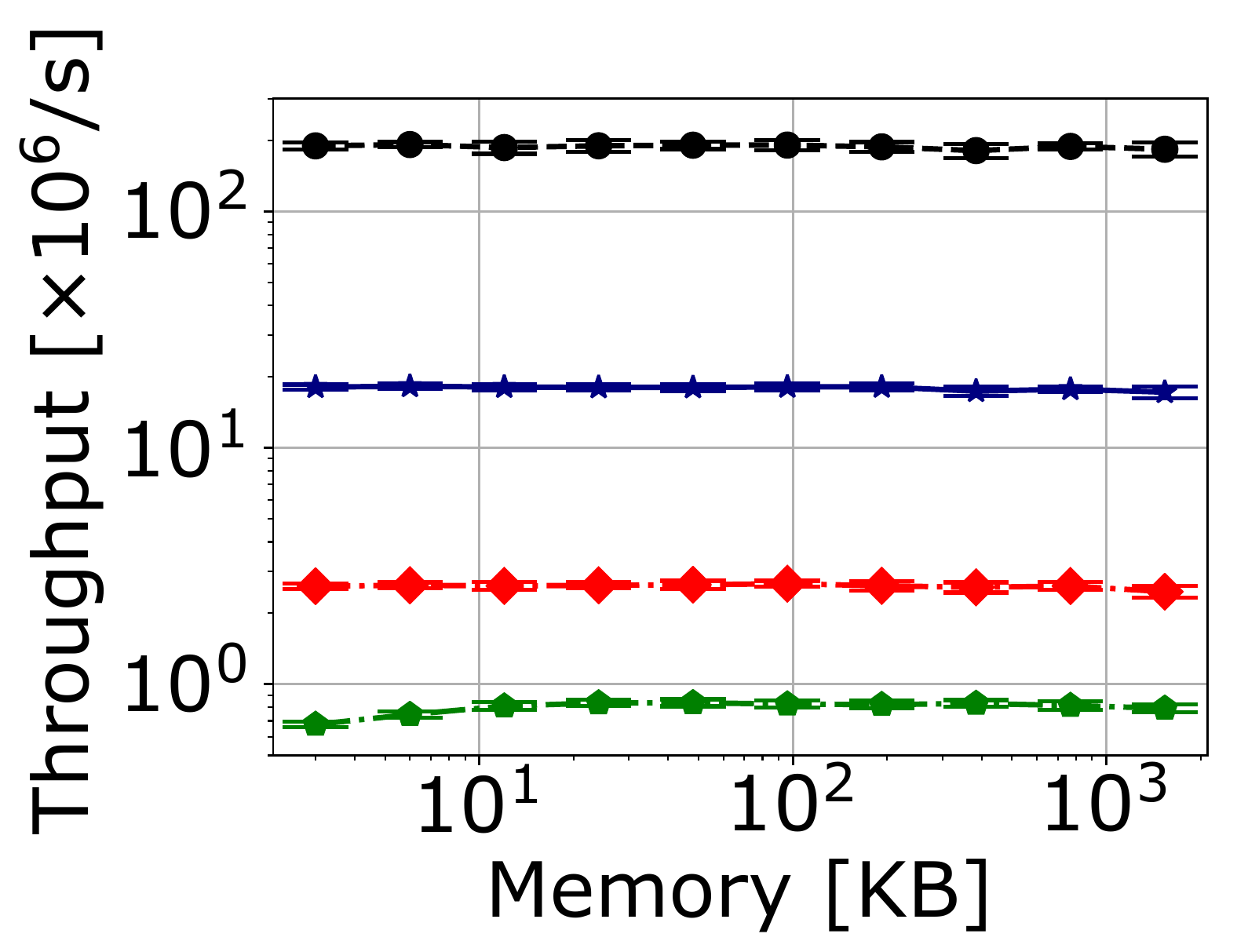}} 
    \vspace*{-1mm}
    \caption{Speed and accuracy of sketch algorithms. All SAC and AEE counters are 16-bits while Baseline uses 32-bits. \label{fig:sketches} }
    \vspace*{-10mm}
\end{figure*}
We use the following metrics; for speed, we use Million operations per second (Mops). For accuracy, on single-estimator experiments, we use Normalized Error, which is defined as the absolute error divided by the number of increments (or the sum of additions in the weighted experiment).

Finally, we run every data point 10 times and use \mbox{Student t-test~\cite{student1908probable} to report the 95\% confidence intervals. }

\subsection{Single Estimator}
We begin by estimating the error and throughput of a single estimator as a function of the number of increments. We compare our Additive Error Estimator (AEE) to Static SAC~\cite{Infocom2019}  and Dynamic SAC~\cite{Infocom2019}. Figure~\ref{2a} shows the normalized error for each 8-bit estimator as a function of the number of increments. 
AEE retains roughly the same normalized error regardless of the number of increments.
In contrast, Static SAC and Dynamic SAC experience higher error and can only count until about $10^3$. 
\MM{I'm going to suggest we remove the 8-bit graphs and just describe verbally that SAC fails with that few bits.  Figure 2 is pretty unreadable;  whatever is going on in 2a, the blowup box within the chart, is way too difficult for to understand.  We don't want to ask reviewers to decipher charts.  We could just show the 16 bit...}\ran{I like the idea of 8-bit estimators. SAC did discuss it a lot (although their estimator is really bad with 8 bits). We can describe in the text the embedded figure in 2(a). Gil/Shay, how do you feel about it?}\ran{we may actually need to cut it for space reasons}
This is because each SAC counter requires few bits to encode its sampling probability, which leaves very few bits for the estimator itself. 
In contrast, all the AEE estimators use the same sampling probability, which means that we can leverage all 8 bits. Figure~\ref{2b} shows the speed of an 8-bit estimator. AEE is orders of magnitude faster since we do not need to access it to decides whether to increment.  
Figure~\ref{2c} and Figure~\ref{2d} repeat this experiment for a 16 bit counter. Static SAC and Dynamic SAC perform better than in the 8-bit case but eventually experience increasing error when the count becomes sufficiently large. In comparison, AEE's error remains the same regardless of the number of increments and is always lower (or equal) to that of Static SAC and Dynamic SAC. Figure~\ref{2d} compares the speed, 
showing that AEE is considerably faster. The non-monotone shape of the AEE curve is due to the computationally expensive random numbers generation. Specifically, AEE is especially fast when not sampling (less than $2^{16}$ increments) and when sampling aggressively (when $N$ is large, and $p$ is small). In between, there is a range in which sampling occurs with a relatively high probability (e.g., 1/2) slowing AEE down. 

\subsection{Sketch Algorithms}
Next, we evaluate the accuracy and speed of the CM sketch~\cite{CountMinSketch} and the CU Sketch~\cite{CUSketch}, using standard 32-bit counters (denoted Baseline), AEE, Dynamic SAC, and Static SAC estimators. Let us first consider the error in the NY18 trace (Figure~\ref{3a} and Figure~\ref{3e}). 
All estimators attain a similar accuracy, which is better than Baseline for both CM Sketch and CU Sketch. Then, as the Memory increases, the precision of the estimator based sketches stops improving while that of the Baseline improves further. 
Intuitively, the error of estimator based sketches has two components. 
One is the sketch error that decreases as we allocate more estimators to the sketch. Another comes from the estimator error that stays the same. Thus, as we gradually reduce the sketch error, it eventually becomes negligible compared to the estimation error. Since the CU Sketch is more accurate than the CM Sketch \cite{CUSketch}, the estimation error becomes the bottleneck earlier. Figure~\ref{3b} and Figure~\ref{3f} repeats this experiment on the CH16 trace.  
The main difference is that the CH16 trace contains only 2.5M distinct flows, while the NY18 trace contains 6.4M distinct flows. As such, the sketch error is considerably lower in the CH16 trace (as there are fewer flows that receive the same counters). Indeed, we see that the error of estimator based sketches does not improve, which implies that estimation error is the dominant one throughout the range. Notably, AEE attains lower error than Static SAC and Dynamic SAC. Figure~\ref{3c}, Figure~\ref{3d}, Figure~\ref{3g} and Figure~\ref{3h} show the speed for the CM Sketch and the CU Sketch.  Static SAC and Dynamic SAC are slower than Baseline because their sampling probability depends on the specific counter. Therefore, for each increment, we first access the sketch counters (and calculate multiple hash functions), and only then determine the sampling probability.
In contrast, in AEE, the sampling probability is identical for all counters. Thus, we first flip a coin and access the sketch counters only if we need to update them. As a result, AEE is \mbox{considerably} faster than Baseline. 





\subsection{Cache-based Algorithms}
\begin{figure}[t]
    \centering
    \subfloat[Error, NY18]
    {\label{4c}\includegraphics[width =0.25\textwidth]
    {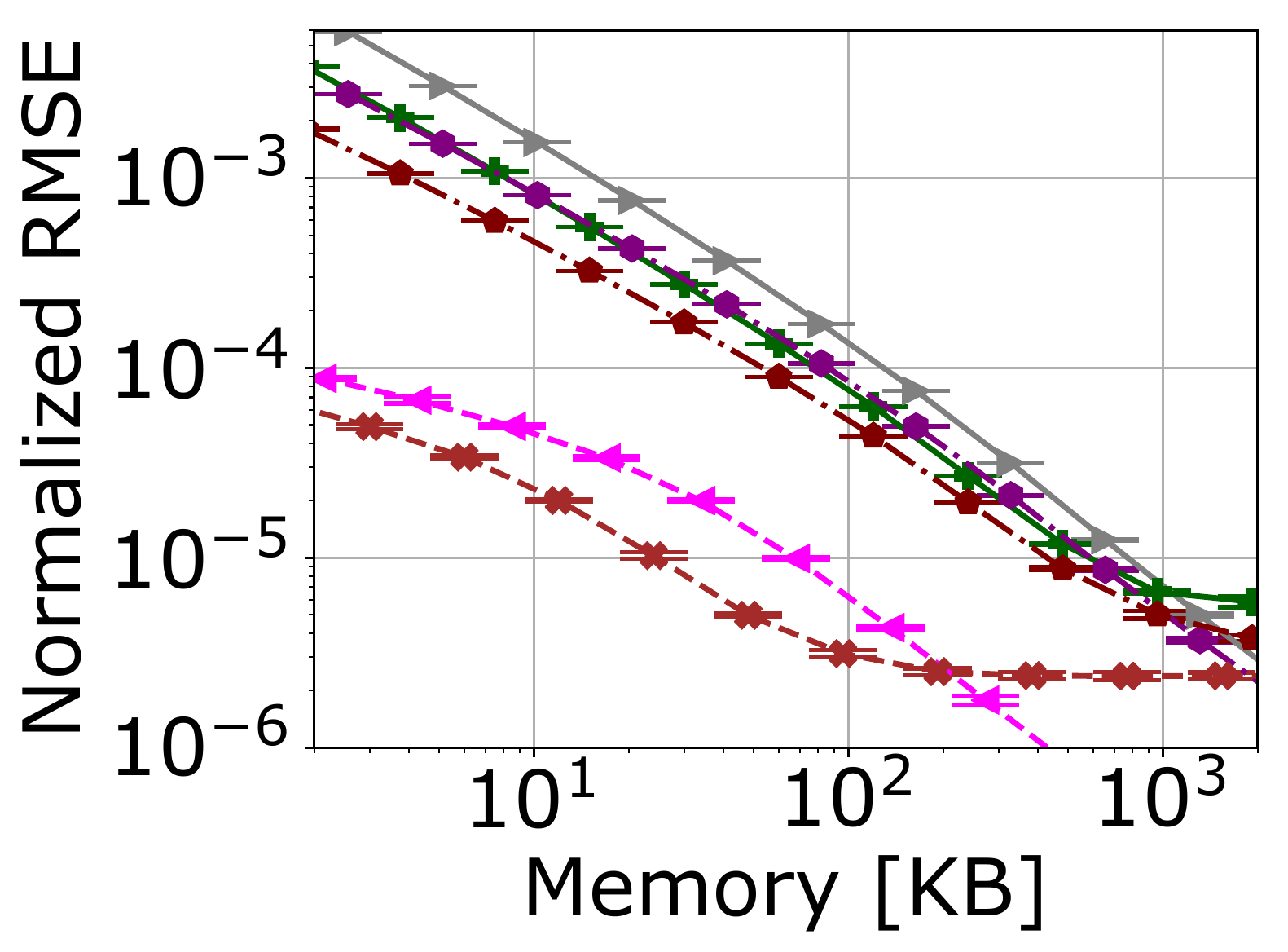}}    
    \subfloat[Speed, NY18]
    {\label{4a}\includegraphics[width =0.25\textwidth]
    {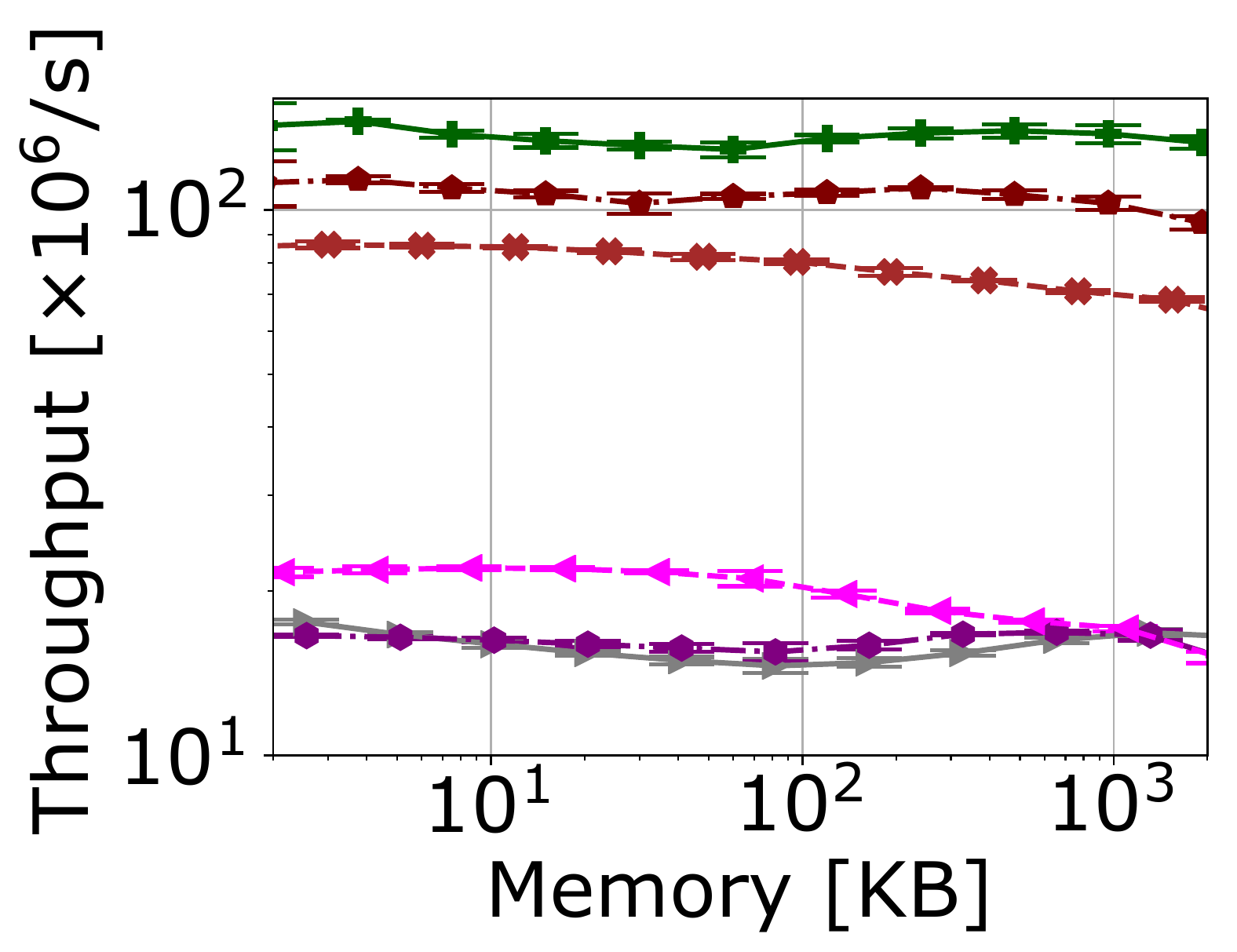}}
\\  
    \hspace*{2mm}{\includegraphics[width =1.0\columnwidth]
    {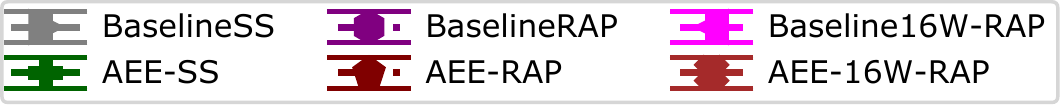}\vspace*{-1mm}}
    \caption{A comparison of cache-based algorithms. 
    \label{fig:cache-based}}\vspace*{-3mm}
\end{figure}

We evaluate our cache-based algorithms compared to their vanilla baseline. Specifically, we compare Space Saving in the original implementation by~\cite{SpaceSavingIsTheBest} (denoted BaselineSS), 
RAP and 16-Way RAP (denoted BaselineRAP and Baseline16W-RAP), and our compressed versions of these algorithms (denoted AAE-SS, AEE-RAP, and AEE-16W-RAP respectively). 
Figure~\ref{4a} shows the update speed. AEE algorithms are an order of magnitude faster than the Baseline algorithms as we do not {need to update the data structures for each packet.}

Figure~\ref{4c}
depicts the error for the NY18 trace. 
At the beginning of the range, each AEE algorithm is more accurate than its corresponding Baseline, and the most accurate ones are Baseline16W-RAP and AEE-16W-RAP. 
At first glance, it may seem strange that we gain better accuracy in the limited associativity model than in the fully associative model. 
However, 16W-RAP can be implemented efficiently in an array, whereas RAP uses the same heap data structure as in the Space Saving implementation,
which requires about 41 bytes per entry~\cite{CormodeCode}. In contrast, 16W-RAP only takes 13 bytes for flow identifier and 4 bytes for the estimator, or a total of 17 bytes per entry. AEE-16W-RAP takes it one step further with just 4 bytes for a fingerprint and 2 for the estimator, i.e., six bytes per entry overall.
Thus, for a given space, Baseline16W-RAP 
has more entries than BaselineRAP, and AEE-16W-RAP has even more. 
As we increase the amount of space, all Baseline algorithms improve, while the AEE algorithms improve until the estimation error becomes the dominant one. 


\subsection{Weighted Counters}
\begin{figure}[t]
    \centering
    \subfloat[16-bit estimators, Error, NY18]
    {\label{5a} \includegraphics[width =0.25\textwidth]
    {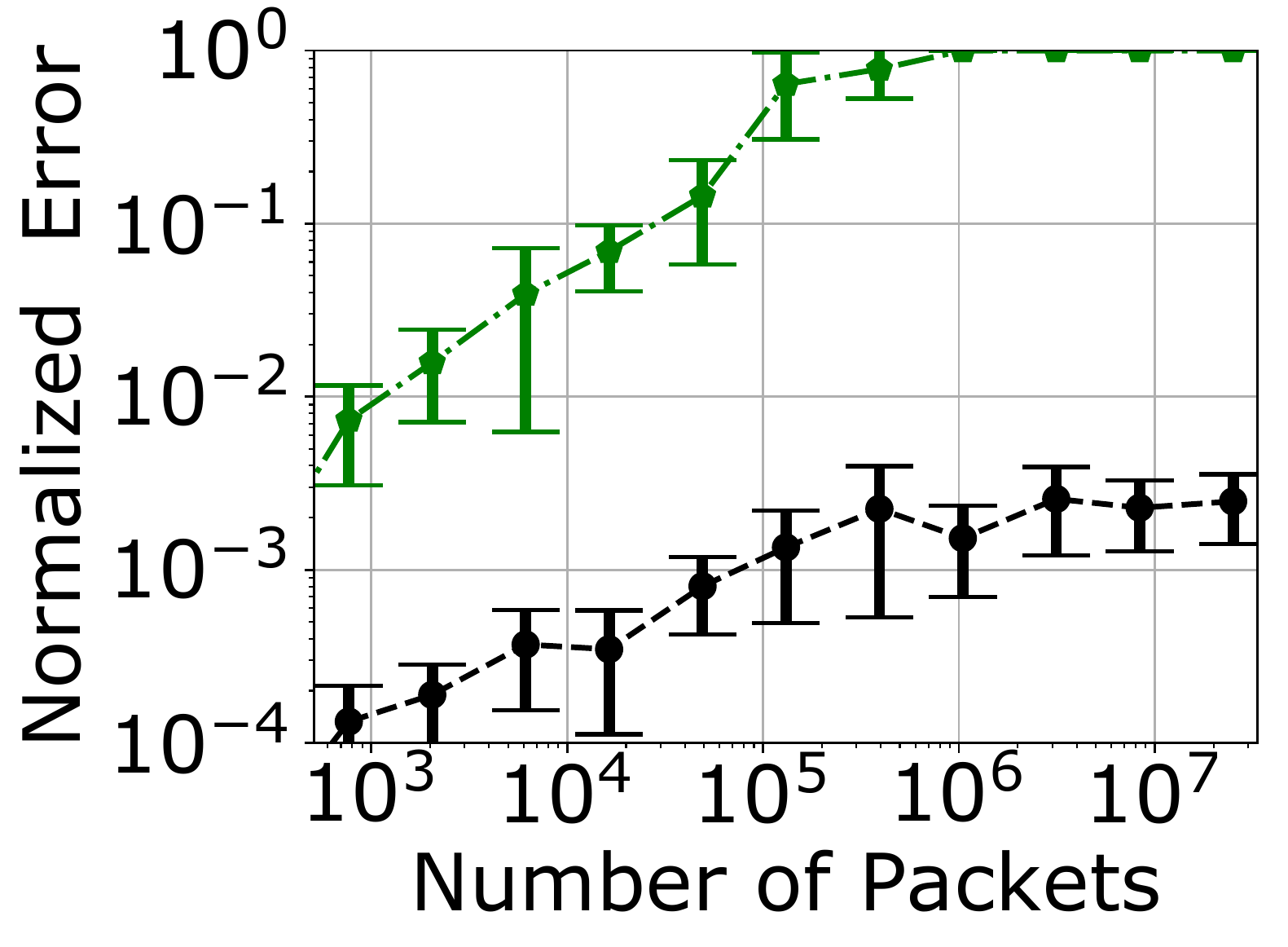}}
    \subfloat[16-bit estimators, Speed, NY18]
    {\label{5b}\includegraphics[width =0.25\textwidth]
    {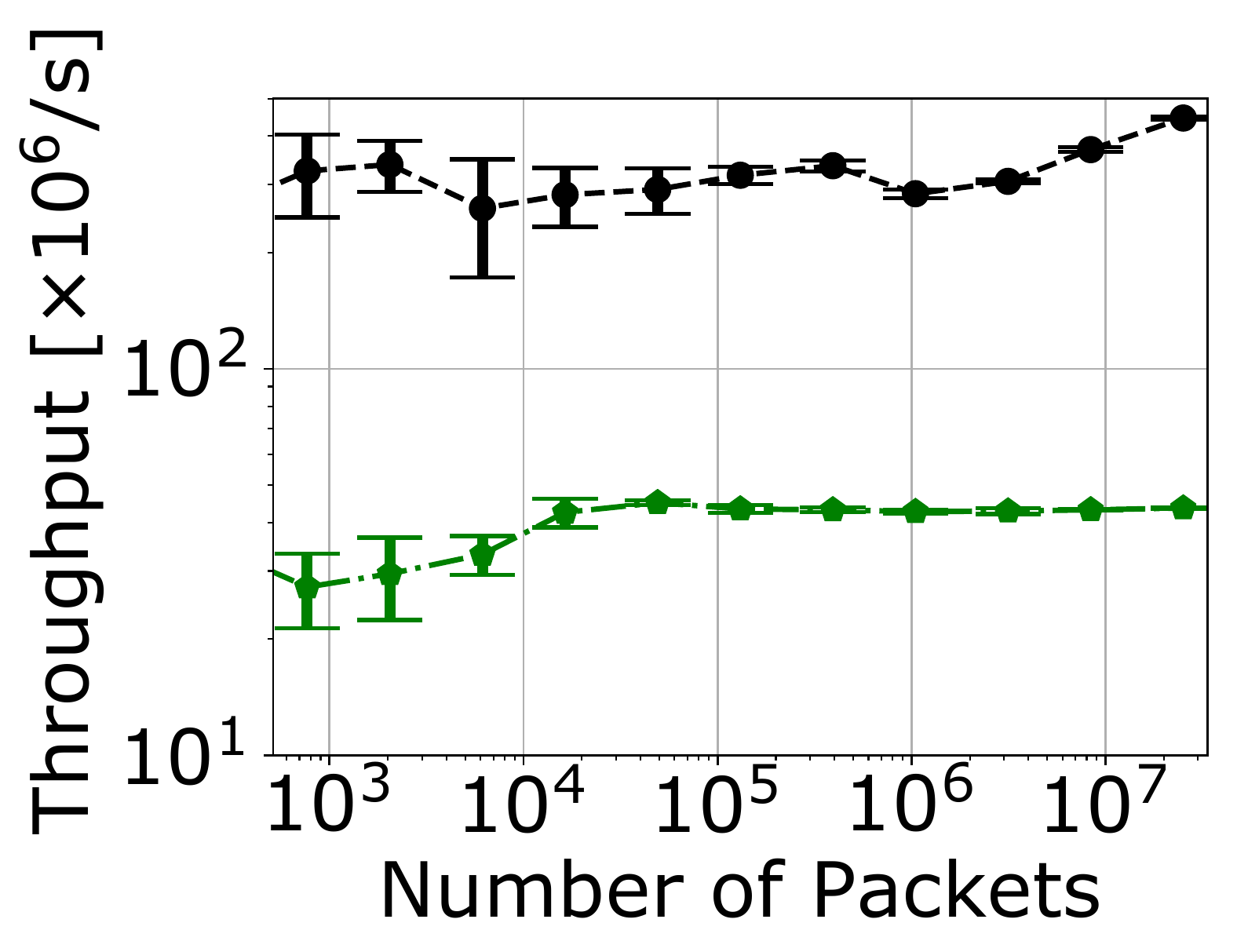}}\\
    {\includegraphics[width =1.04\columnwidth]
    {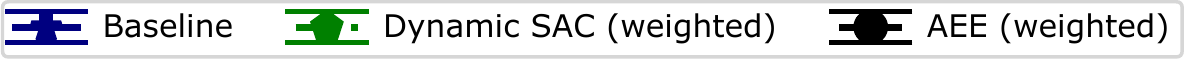}\vspace*{-3mm}}
    \subfloat[Weighted CM Sketch, Error, NY18]
    {\label{5c} \includegraphics[width =0.25\textwidth]
    {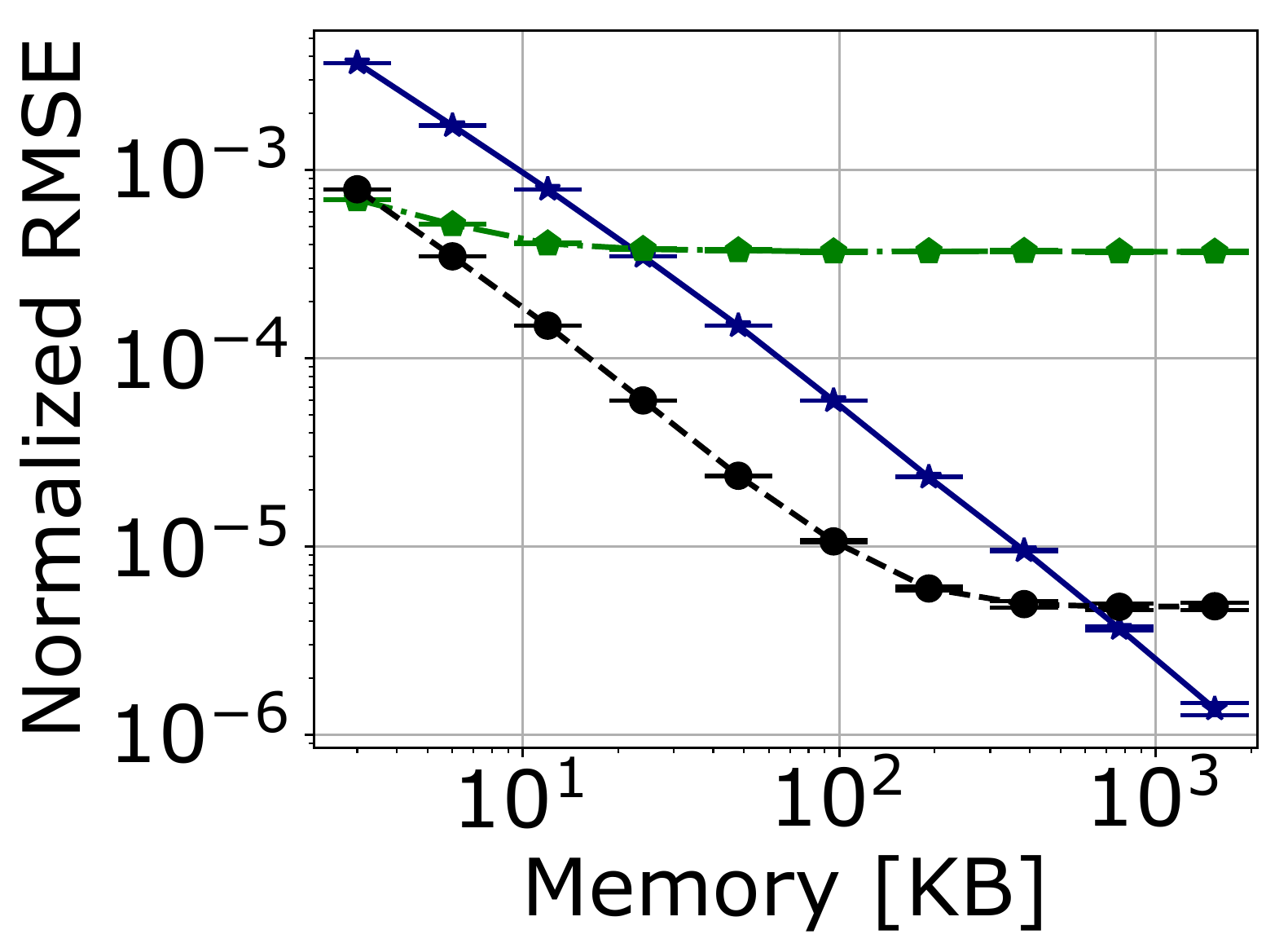}}
    \subfloat[Weighted CM Sketch, Speed, NY18]
    {\label{5d}\includegraphics[width =0.25\textwidth]
    {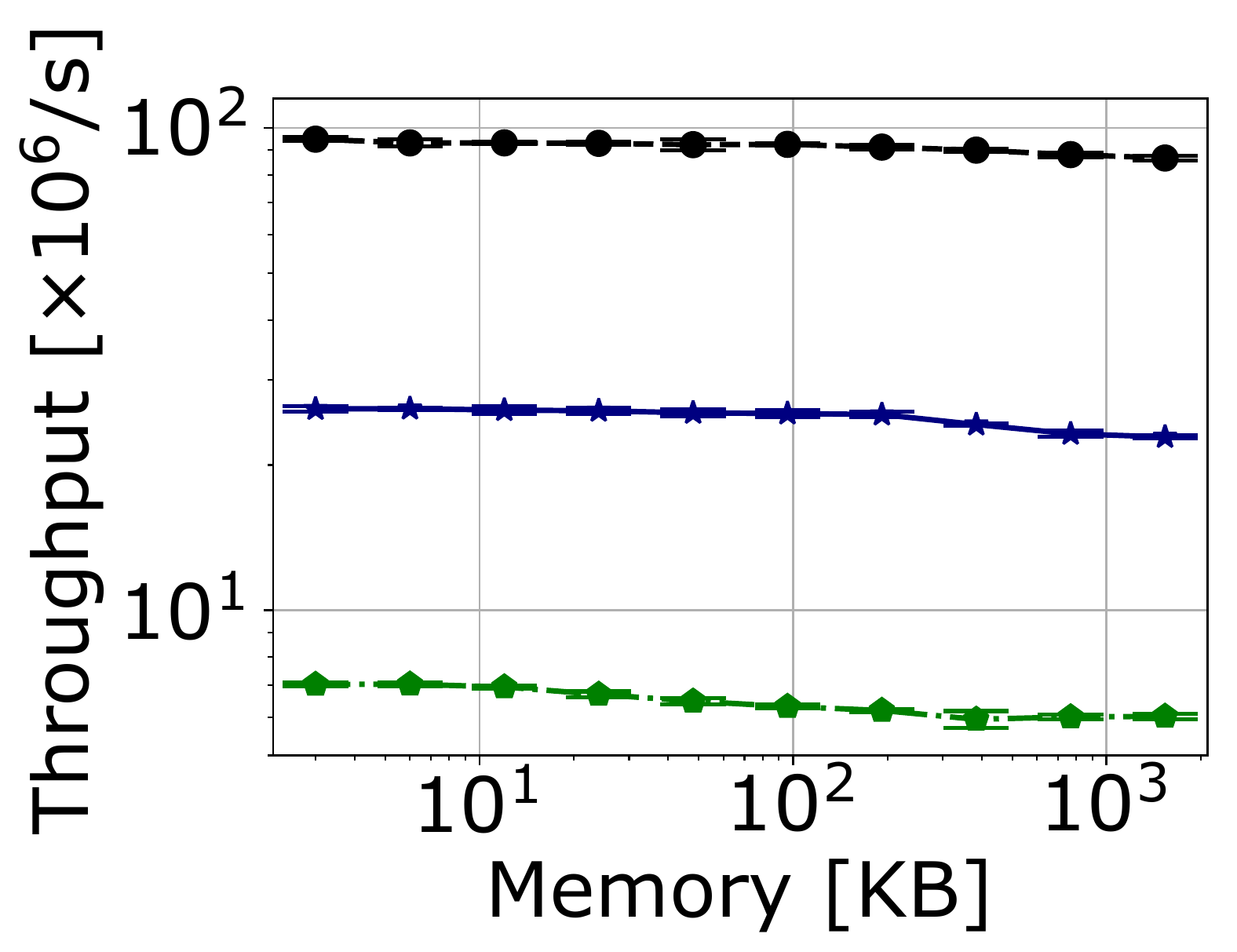}} 
    \vspace*{-1mm}
    \caption{\small A comparison of the speed and accuracy of single weighted estimators and weighted Count Min Sketch (NY18 trace). The SAC and AEE estimators are 16-bits while the Baseline uses 64.}\label{fig:weighted}
    \vspace*{-3mm}
\end{figure}

We estimate the total byte volume of the NY18 trace using a single estimator. 
The results are depicted in Figures~\ref{5a} and~\ref{5b}. As in the unweighted case, AEE has better accuracy (${\approx}100\times$) and speed (${\approx}8\times$) compared with Dynamic SAC. 

Figures~\ref{5c} and~\ref{5d} show results for per-flow byte volume estimation on the NY18 trace. 
AEE is more accurate than the baseline ($\approx 7\times$) until the estimation error becomes dominant ($\approx 800KB$). AEE is also faster than the Baseline ($\approx 4.5\times$). 
For accuracy, Dynamic SAC shows a similar trend, but its estimation error becomes dominant at a smaller size. 

\vspace*{-1mm}
\subsection{The {\sc MaxSpeed} Variant}\label{sec:maxspeedeval}
\begin{figure}[t]
    \centering
    \subfloat[CM Sketch, Error, NY18]
    {\label{6a} \includegraphics[width =0.25\textwidth]
    {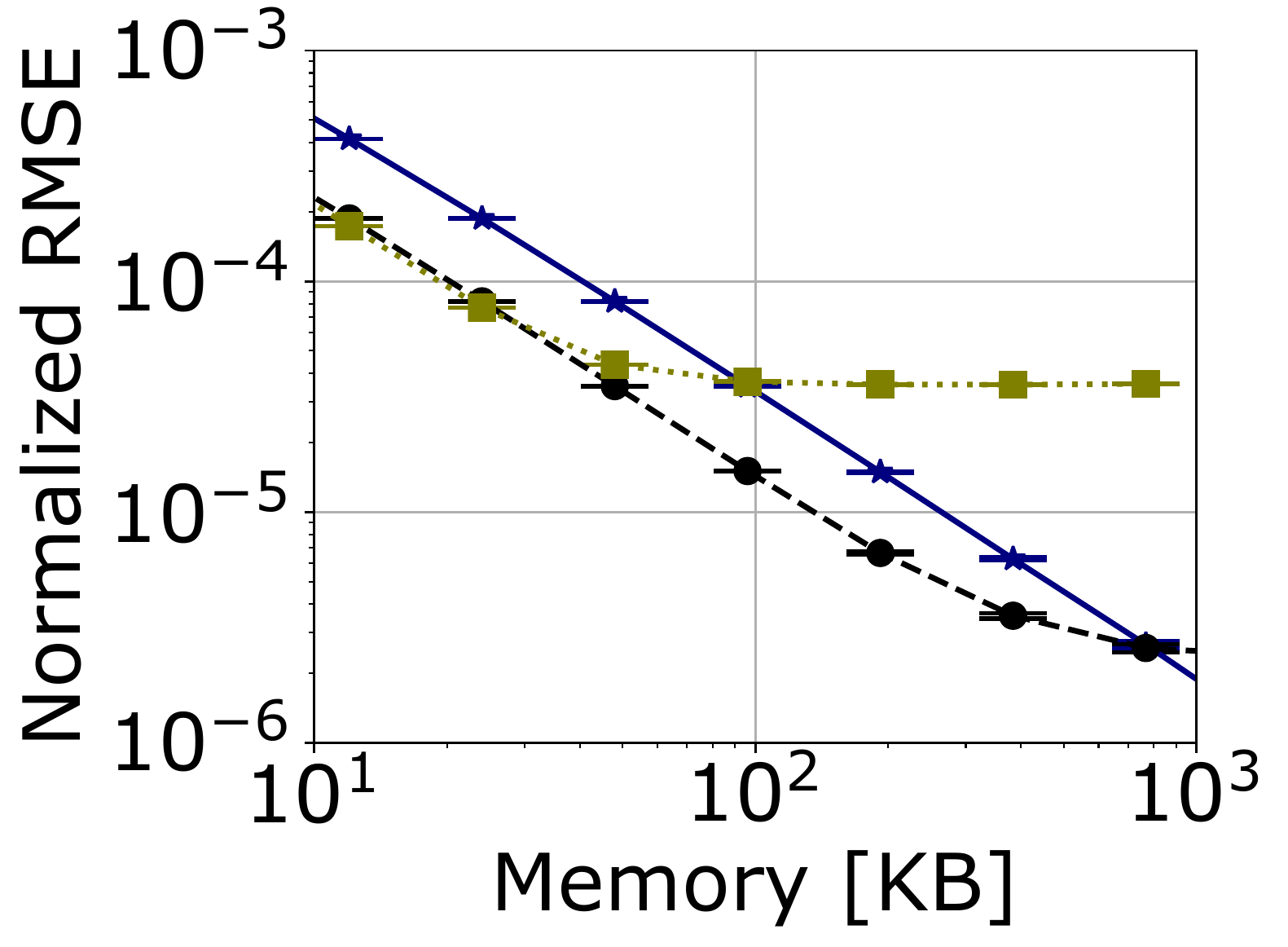}}
    \subfloat[CM Sketch, Speed, NY18]
    {\label{6b}\includegraphics[width =0.25\textwidth]
    {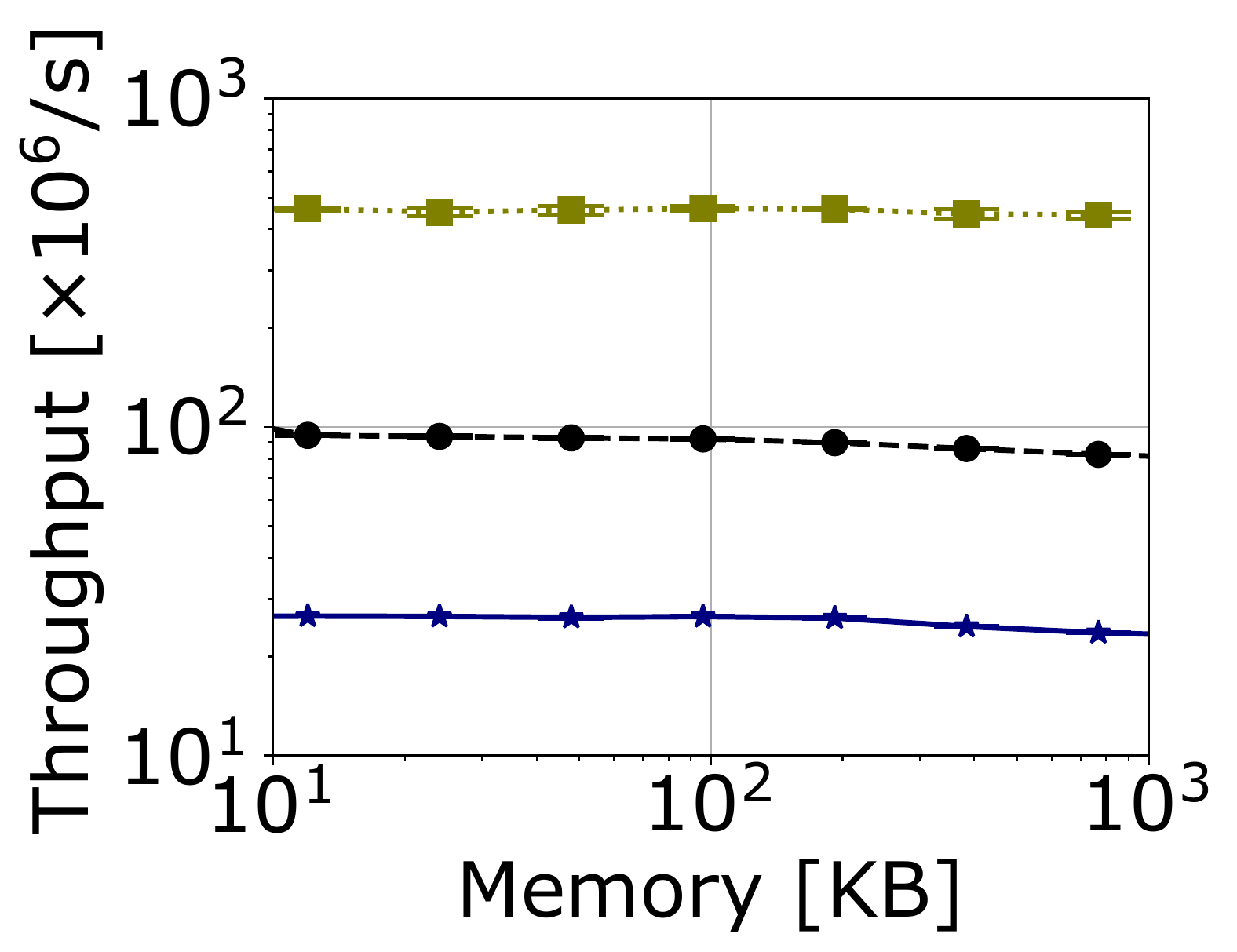}}\\
    {\includegraphics[width =1.02\columnwidth]
    {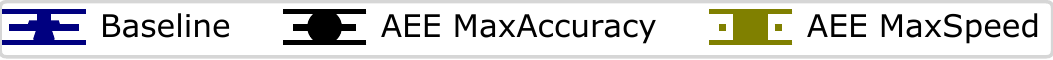}}
    \vspace*{-3mm}
    \caption{Comparing the {\sc MaxAccuracy} and {\sc MaxSpeed} variants of AEE \mbox{on Count Min Sketch and NY18 data.}}\label{fig:maxspeed} 
    \vspace*{-4mm}
\end{figure}

We now evaluate {\sc MaxSpeed} versus {\sc MaxAccuracy} (which we used in previous sections). 
As shown in Figure~\ref{fig:maxspeed}, {\sc MaxSpeed} is about $4\times$ faster than {\sc MaxAccuracy} while offering similar accuracy when the allocated memory is small. We conclude that {\sc MaxSpeed} is suitable when space is tight or if one requires extremely high speeds.
\section{Discussion}
Our work explores the opportunities offered by replacing full-sized counters in approximate measurement algorithms with short estimators. Specifically, we observe that the target algorithms provide additive error guarantees, while most estimators are designed to provide multiplicative error, which adds needless complexity in this context. 

We introduce an Additive Error Estimator (AEE) that offers benefits over multiplicative estimators when combined with sketches and cache-based counting algorithms. Most notably, it maintains the same $N\epsilon$ additive error guarantee over any counting range.  Namely, AEE allows us to count indefinitely without overflowing while maintaining the accuracy guarantee.  
Further, AEE offers faster update speed as it increments all counters with the same probability and avoids computing hash functions for non-sampled packets. 
%
Our empirical results show that the AEE estimator is faster and more accurate than existing estimators.  
The evaluation also shows the limitations of our estimator, which are in line with the theoretical results.   

The code of our algorithms is available as open source~\cite{opensource}.
\vspace*{-1mm}

\vspace*{-1mm}
\appendix
\vspace*{-1mm}
\subsection{Proof of our single counter correctness}\label{app:singleCounterProof}
\single
\begin{proof}
If the number of {\sc Increment}s was $I$, then $C\sim\mbox{Bin}(I,p)$. We have that $\mathbb E[C]=Ip$ and $\Var[C]=Ip(1-p)$.
We use a variant of the Bennett bound (see~\cite[Eq.1.15]{janson2016large}) stating that for every set of $\set{X_i}$ independent Bernoulli random variables such that $X_i=1 \text{ w.p. } p_i$ and $0$ otherwise, 
their sum $X{=}\sum_{i=1}^n X_i$ satisfies $\forall a,z{>}0$ such that $a{\ge}z\Var[X]$:
$$
{
\vspace*{-1mm}
\Pr[|X-\mathbb E[X]|\ge a] \le e^{-a((1-1/z)\ln(1+z)-1)}.
}
$$
Consider our counter $C=\sum_{i=1}^I X_i$ where $X_i$ is the indicator of the event in which the $i$'th attempted increment operation increased the counter.
Choosing $a=\frac{z\Var[C]}{1-p}=z\mathbb E[C]$ we get that for all $z>0$:
\begin{equation}\label{eq:Bennet}
\vspace*{-2mm}
     \hspace{-.5mm}\Pr\brackets{|C-\mathbb E[C]|\ge z\mathbb E[C]}\le 2e^{-z\mathbb E[C]\parentheses{(1+1/z)\ln(1+z)-1}}.\hspace{-.5mm}
\end{equation}
We use~\eqref{eq:Bennet} for our counter $C$,  
and set $z=\frac{N\epsilon}{I}$ to obtain: 
\begin{multline*}
    \Pr[|C/p-I|> N\epsilon] = \Pr[|C-Ip|> Np\epsilon] 
    \\= \Pr[|C-\mathbb E[C]|> zIp]
    = \Pr[|C-\mathbb E[C]|> z\mathbb E[C]]
    \\\le e^{-z\mathbb E[C]\parentheses{(1+1/z)\ln(1+z)-1}}
    = 2e^{-\mathbb E[C]\parentheses{(1+z)\ln(1+z)-z}}
    \\
    = 2e^{-Ip\parentheses{(1+\frac{N\epsilon}{I})\ln(1+\frac{N\epsilon}{I})-\frac{N\epsilon}{I}}}
    = 2e^{-p\parentheses{(I+N\epsilon)\ln(1+\frac{N\epsilon}{I})-N\epsilon}}\\
    = 2e^{N\epsilon p}\cdot \parentheses{(1+{N\epsilon}/{I})^{-\parentheses{I+N\epsilon}}}^p.
    \vspace*{-2mm}
\end{multline*}
The function $(1+{N\epsilon}/{I})^{-\parentheses{I+N\epsilon}}$ is monotonically increasing in $I$, and therefore so is $2e^{-p\parentheses{(I+N\epsilon)\ln(1+\frac{N\epsilon}{I})-N\epsilon}}$. As $I\le N$ we can bound the error probability as 
{\small
\begin{multline*}
\vspace*{-2mm}
    \Pr[|C/p-I|> N\epsilon]\le e^{-p\parentheses{(I+N\epsilon)\ln(1+\frac{N\epsilon}{I})-N\epsilon}}\\
    \le 2e^{-p\parentheses{(N+N\epsilon)\ln(1+\frac{N\epsilon}{N})-N\epsilon}}
    = 2e^{-Np\parentheses{(1+\epsilon)\ln(1+\epsilon)-\epsilon}}.
    \vspace*{-2mm}
\end{multline*}
}
We use the elementary inequality {\small$(1+\epsilon)\ln(1+\epsilon)-\epsilon \ge \frac{\epsilon^2}{2(1+\epsilon/3)},$} which gets us to 
$$
    \Pr[|C/p-I|> N\epsilon]
    \le 2e^{-Np\epsilon^{2}/2\parentheses{1+\epsilon/3}}\le \delta,
$$
where the last inequality \mbox{follows from our choice of $p$.}
\vspace*{-2mm}
\end{proof}
\subsection{Proof of our weighted updates correctness}\label{app:weightedUpdatesProof}
Consider a stream of weighted updates $\mathfrak w_{(1)},\ldots,\mathfrak w_{(q)}$ and let $W=\sum_{i=1}^q \mathfrak w_{(i)}$ denote the total additions made to the counter.
For each $i$, let $w_{1,(i)}=\floor{\mathfrak w_{(i)}\cdot p}$ and $w_{2,(i)}=\mathfrak w-w_{1,(i)}$ denote the partitioning of the weight as explained in Section~\ref{sec:weighted}. We also use $W_1=\sum_{i=1}^q \mathfrak w_{1,(i)}$ and  $W_2=\sum_{i=1}^q \mathfrak w_{2,(i)}$ to denote the partial weights.

For each $i\in\set{1,\ldots,q}$, let $X_i$ denote whether we incremented the counter as a result of the coin flip for the $i$'th update, i.e.,
$X_i = 1 \text{ w.p. } w_{2,(i)}\cdot p$ and $0$ otherwise.
Observe that $C=\sum_{i=1}^q w_{1,(i)}/p + X_i$.
As the first summand is deterministic, we denote $\widetilde C = \sum_{i=1}^q X_i$ for its probabilistic part; we have that $C/p= W_1 + \widetilde C/p$ and $\mathbb E[\widetilde C]=W_2 p$.
Our goal is to show that $\Var[\widetilde C]\le W_2 p (1-p)$ as this would imply the correctness of our algorithm \mbox{similarly to the unweighted case:}
{\small\vspace*{-2mm}
\begin{multline*}
 \Var[\widetilde C] = \sum_{i=1}^q\Var[X_i] = \sum_{i=1}^q w_{2,(i)} p (1-w_{2,(i)}\cdot p)\\
\le \sum_{i=1}^q w_{2,(i)} p (1-p) = W_2 p (1-p).
\vspace*{-2mm}
\end{multline*}}
That is, we showed that $\mathbb E[C]=W$ and that $\Var[C]=\Var[\widetilde C] \le W_2 p (1-p)\le Wp(1-p)$.
The correctness then follows from an analysis similar to that of Appendix~\ref{app:singleCounterProof}.

\subsection{Proof of the sum-of-counters Bound}\label{app:sumOfCountersBoundProof}
We now prove that the sum of compressed counters in our counter array is at most $\widetilde{N'}\triangleq N'+\sqrt{3N'\ln\delta_o^{-1}}$ with probability $1-\delta_o$.
Let $I_i$ the number of times an {\sc Increment}$(i)$ operation was called, for $i\in\set{1,\ldots,w}$, and let $I=\sum_{i=1}^w I_i$ denote the total number of increments. 
Notice that since $I\le N$ we have $\mathbb E[I]\le N'$. For $j\in\set{1,\ldots,I}$, let $X_j$ denote whether the $j$'th increment operation (to any counter) resulted in an increase in a counter.
We denote by $X=\sum_{j=1}^I X_j$ the sum of all counters after the $I$ increments. Then $X\sim\mbox{Bin}(I,p)$ and a simple application of the Chernoff bound implies $\Pr[X\ge \widetilde{N'}]\le \delta_o$.

\newpage

{ 
	\bibliographystyle{IEEEtran}
	\bibliography{references}

\begin{thebibliography}{10}
\providecommand{\url}[1]{#1}
\csname url@samestyle\endcsname
\providecommand{\newblock}{\relax}
\providecommand{\bibinfo}[2]{#2}
\providecommand{\BIBentrySTDinterwordspacing}{\spaceskip=0pt\relax}
\providecommand{\BIBentryALTinterwordstretchfactor}{4}
\providecommand{\BIBentryALTinterwordspacing}{\spaceskip=\fontdimen2\font plus
\BIBentryALTinterwordstretchfactor\fontdimen3\font minus
  \fontdimen4\font\relax}
\providecommand{\BIBforeignlanguage}[2]{{%
\expandafter\ifx\csname l@#1\endcsname\relax
\typeout{** WARNING: IEEEtran.bst: No hyphenation pattern has been}%
\typeout{** loaded for the language `#1'. Using the pattern for}%
\typeout{** the default language instead.}%
\else
\language=\csname l@#1\endcsname
\fi
#2}}
\providecommand{\BIBdecl}{\relax}
\BIBdecl

\bibitem{LoadBalancing}
G.~Dittmann and A.~Herkersdorf, ``Network processor load balancing for
  high-speed links,'' in \emph{SPECTS}, 2002.

\bibitem{TrafficEngeneering}
T.~Benson, A.~Anand, A.~Akella, and M.~Zhang, ``Microte: Fine grained traffic
  engineering for data centers,'' in \emph{ACM CoNEXT}, 2011.

\bibitem{SLA}
J.~Sommers, P.~Barford, N.~Duffield, and A.~Ron, ``Accurate and efficient sla
  compliance monitoring,'' ser. ACM SIGCOMM, 2007.

\bibitem{IntrusionDetection}
B.~Mukherjee, L.~Heberlein, and K.~Levitt, ``Network intrusion detection,''
  \emph{Network, IEEE}, 1994.

\bibitem{IntrusionDetection2}
P.~Garcia-Teodoro, J.~E. Díaz-Verdejo, G.~Maciá-Fernández, and E.~Vázquez,
  ``Anomaly-based network intrusion detection: Techniques, systems and
  challenges,'' \emph{Computers and Security}, 2009.

\bibitem{Nitro}
Z.~Liu, R.~Ben-Basat, G.~Einziger, Y.~Kassner, V.~Braverman, R.~Friedman, and
  V.~Sekar, ``Nitrosketch: Robust and general sketch-based monitoring in
  software switches,'' in \emph{ACM SIGCOMM}, 2019.

\bibitem{RHHH}
R.~Ben~Basat, G.~Einziger, R.~Friedman, M.~C. Luizelli, and E.~Waisbard,
  ``Constant time updates in hierarchical heavy hitters,'' in \emph{ACM
  SIGCOMM}, 2017.

\bibitem{Brick}
N.~Hua, B.~Lin, J.~J. Xu, and H.~C. Zhao, ``Brick: A novel exact active
  statistics counter architecture,'' in \emph{ACM/IEEE ANCS}, 2008.

\bibitem{univmon}
Z.~Liu, A.~Manousis, G.~Vorsanger, V.~Sekar, and V.~Braverman, ``One sketch to
  rule them all: Rethinking network flow monitoring with univmon,'' in
  \emph{ACM SIGCOMM}, 2016.

\bibitem{CountSketch}
M.~Charikar, K.~Chen, and M.~Farach-Colton, ``Finding frequent items in data
  streams,'' in \emph{EATCS ICALP}, 2002.

\bibitem{CountMinSketch}
G.~Cormode and S.~Muthukrishnan, ``An improved data stream summary: The
  count-min sketch and its applications,'' \emph{J. Algorithms}, 2004.

\bibitem{RandomizedCounterSharing}
T.~Li, S.~Chen, and Y.~Ling, ``Per-flow traffic measurement through randomized
  counter sharing,'' \emph{IEEE/ACM Trans. on Networking}, 2012.

\bibitem{SketchVisor}
Q.~Huang, X.~Jin, P.~P.~C. Lee, R.~Li, L.~Tang, Y.-C. Chen, and G.~Zhang,
  ``Sketchvisor: Robust network measurement for software packet processing,''
  in \emph{ACM SIGCOMM}, 2017.

\bibitem{SpaceSavings}
A.~Metwally, D.~Agrawal, and A.~E. Abbadi, ``Efficient computation of frequent
  and top-k elements in data streams,'' in \emph{ICDT}, 2005.

\bibitem{SAC}
R.~Stanojevic, ``Small active counters,'' in \emph{IEEE INFOCOM}, 2007.

\bibitem{DISCO}
C.~Hu, B.~Liu, H.~Zhao, K.~Chen, Y.~Chen, C.~Wu, and Y.~Cheng, ``Disco: Memory
  efficient and accurate flow statistics for network measurement,'' in
  \emph{IEEE ICDCS}, 2010.

\bibitem{CEDAR}
E.~Tsidon, I.~Hanniel, and I.~Keslassy, ``Estimators also need shared values to
  grow together,'' in \emph{IEEE INFOCOM}, 2012.

\bibitem{ICE-Buckets}
G.~{Einziger}, B.~{Fellman}, R.~{Friedman}, and Y.~{Kassner}, ``Ice buckets:
  Improved counter estimation for network measurement,'' \emph{IEEE/ACM
  Transactions on Networking}, 2018.

\bibitem{CASE}
L.~Yang, W.~Hao, P.~Tian, D.~Huichen, L.~Jianyuan, and L.~Bin, ``Case:
  Cache-assisted stretchable estimator for high speed per-flow measurement,''
  in \emph{IEEE INFOCOM}, 2016.

\bibitem{Infocom2019}
T.~{Yang}, J.~{Xu}, X.~{Liu}, P.~{Liu}, L.~{Wang}, J.~{Bi}, and X.~{Li}, ``A
  generic technique for sketches to adapt to different counting ranges,'' in
  \emph{IEEE INFOCOM}, 2019.

\bibitem{CUSketch}
C.~Estan and G.~Varghese, ``New directions in traffic measurement and
  accounting,'' \emph{ACM SIGCOMM}, 2002.

\bibitem{ApproximateCounting}
R.~Morris, ``Counting large numbers of events in small registers,''
  \emph{Commun. ACM}, 1978.

\bibitem{ANLS}
C.~Hu, S.~Wang, B.~L. Tian, Jia, Y.~Cheng, and Y.~Chen, ``Accurate and
  efficient traffic monitoring using adaptive non-linear sampling method,'' in
  \emph{IEEE INFOCOM}, 2008.

\bibitem{ANLSUpscaling}
C.~Hu and B.~Liu, ``Self-tuning the parameter of adaptive non-linear sampling
  method for flow statistics,'' in \emph{CSE}, 2009.

\bibitem{SpectralBloom}
S.~Cohen and Y.~Matias, ``Spectral bloom filters,'' in \emph{ACM SIGMOD}, 2003.

\bibitem{CounterBraids}
Y.~Lu, A.~Montanari, B.~Prabhakar, S.~Dharmapurikar, and A.~Kabbani, ``Counter
  braids: a novel counter architecture for per-flow measurement,'' in \emph{ACM
  SIGMETRICS}, 2008.

\bibitem{countertree}
M.~Chen and S.~Chen, ``Counter tree: {A} scalable counter architecture for
  per-flow traffic measurement,'' in \emph{IEEE ICNP}, 2015.

\bibitem{frequent4}
E.~D. Demaine, A.~L\'{o}pez-Ortiz, and J.~I. Munro, ``Frequency estimation of
  internet packet streams with limited space,'' in \emph{EATCS ESA}, 2002.

\bibitem{SpaceSavingIsTheBest}
G.~Cormode and M.~Hadjieleftheriou, ``Finding frequent items in data streams,''
  \emph{VLDB}, 2008.

\bibitem{HashPipe}
V.~Sivaraman, S.~Narayana, O.~Rottenstreich, S.~Muthukrishnan, and J.~Rexford,
  ``Heavy-hitter detection entirely in the data plane,'' in \emph{ACM SOSR},
  2017.

\bibitem{HeavyHitters}
R.~Ben-Basat, G.~Einziger, R.~Friedman, and Y.~Kassner, ``Heavy hitters in
  streams and sliding windows,'' in \emph{IEEE INFOCOM}, 2016.

\bibitem{10.14778/3297753.3297762}
R.~B. Basat, R.~Friedman, and R.~Shahout, ``Stream frequency over interval
  queries,'' in \emph{VLDB}, 2019.

\bibitem{SpaceSavingIsTheBest2010}
G.~C{or}mode and M.~Hadjieleftheriou, ``Methods for finding frequent items in
  data streams,'' \emph{J. VLDB}, 2010.

\bibitem{SpaceSavingIsTheBest2011}
N.~Manerikar and T.~Palpanas, ``Frequent items in streaming data: An
  experimental evaluation of the state-of-the-art,'' \emph{Data Knowl. Eng.},
  2009.

\bibitem{misra1982finding}
J.~Misra and D.~Gries, ``Finding repeated elements,'' \emph{Science of computer
  programming}, 1982.

\bibitem{BatchDecrement}
R.~M. Karp, S.~Shenker, and C.~H. Papadimitriou, ``A simple algorithm for
  finding frequent elements in streams and bags,'' \emph{ACM Trans. Database
  Syst.}, 2003.

\bibitem{RAP}
R.~{Ben-Basat}, X.~{Chen}, G.~{Einziger}, R.~{Friedman}, and Y.~{Kassner},
  ``Randomized admission policy for efficient top-k, frequency, and volume
  estimation,'' \emph{IEEE/ACM Transactions on Networking}, 2019.

\bibitem{dimsum}
R.~Ben{-}Basat, G.~Einziger, R.~Friedman, and Y.~Kassner, ``Optimal elephant
  flow detection,'' in \emph{Proceedings of IEEE Infocom}, 2017.

\bibitem{IMSUM}
D.~Anderson, P.~Bevan, K.~J. Lang, E.~Liberty, L.~Rhodes, and J.~Thaler, ``A
  high-performance algorithm for identifying frequent items in data streams,''
  in \emph{ACM IMC}, 2017.

\bibitem{TinyTable}
G.~Einziger and R.~Friedman, ``Counting with tinytable: Every bit counts!''
  \emph{IEEE Access}, 2019.

\bibitem{TinyTable2}
P.~Pandey, M.~A. Bender, R.~Johnson, and R.~Patro, ``A general-purpose counting
  filter: Making every bit count,'' in \emph{ACM SIGMOD}, 2017.

\bibitem{gibbons1998new}
P.~B. Gibbons and Y.~Matias, ``New sampling-based summary statistics for
  improving approximate query answers,'' in \emph{Sigmod Record}, 1998.

\bibitem{student1908probable}
Student, ``The probable error of a mean,'' \emph{Biometrika}, 1908.

\bibitem{CormodeCode}
\BIBentryALTinterwordspacing
G.~Cormode, ``Implementation of heavy hitter algorithms.'' [Online]. Available:
  \url{http://hadjieleftheriou.com/frequent-items/}
\BIBentrySTDinterwordspacing

\bibitem{CAIDA2018}
``The caida equinix-newyork packet trace, 20181220-130000.''

\bibitem{CAIDA2016}
``The caida equinix-chicago packet trace, 20160406-130000.''

\bibitem{opensource}
``Open source code.'' \url{https://github.com/additivecounters/AEE}.

\bibitem{janson2016large}
S.~Janson, ``Large deviation inequalities for sums of indicator variables,''
  \emph{arXiv preprint arXiv:1609.00533}, 2016.

\end{thebibliography}
}


\end{document}